\documentclass{aa}  
\usepackage[version=3]{mhchem} 
\usepackage{natbib}
\usepackage{graphicx}
\usepackage{subcaption}  
\usepackage[dvipsnames]{xcolor}
\usepackage[normalem]{ulem}
\usepackage{mathrsfs}
\usepackage{verbatim}
\usepackage{cancel}
\usepackage{url}
\usepackage{textgreek}
\usepackage{txfonts}
\def\cje#1{#1}

\begin{document}

   \title{Silicate cosmic dust grain collisions in the interstellar medium}

   \subtitle{A molecular dynamics study}

   \author{C.J. Esmerian\inst{1}\fnmsep\thanks{corresponding author, \email{clarke.esmerian@chalmers.se}}
          \and
          S.R. Hashemi\inst{2}\fnmsep\thanks{The first two authors contributed equally.}
          \and
          W.M.C. Sameera\inst{1}
          \and
          \cje{B. Lankhaar}\inst{3}
          \and 
          W. Vlemmings\inst{1}
          \and
          S. Andersson\inst{2,4}
          \and
          T. J. L. C. Bakx\inst{1}
          \and
           K. K. Knudsen\inst{1}
          \and
          S. Aalto\inst{1}
          \and
          G. Nyman\inst{2}
          }

   \institute{
        Department of Physics and Astronomy, Chalmers University of Technology, SE-412 96 Gothenburg, Sweden
        \and
        University of Gothenburg, Department of Chemistry and Molecular Biology, Box 462, 40530 Gothenburg, Sweden
        \and
        Institute of Theoretical Astrophysics, University of Oslo, P.O. Box 1029, Blindern, 0315 Oslo, Norway
        \and
        SINTEF, P.O. Box 4760 Torgarden, NO-7465 Trondheim, Norway
        }

   \date{}

 
  \abstract
   {Cosmic dust grains are an important component of the interstellar medium (ISM) and observations suggest a distribution of grain sizes from several~\AA{} to $\sim \mu$m. Collisions between grains, which can result in coagulation, shattering, or vaporization, likely influence the grain size distribution.}
   {We aim to predict the most important parameters for grain-grain collision outcomes for models of grain population evolution on astrophysical scales: the threshold velocity above which colliding grains shatter, the threshold for vaporization, and resulting distributions of grain sizes.} 
   {For the first time (to the knowledge of the authors), we use molecular dynamics simulations which evolve the dynamics of each atom in a dust grain to explore the outcomes of collisions between silicate grains of radii $a \in [5,50]$~\AA{} at velocities expected for these grains in the interstellar medium, $0.1-20$ km/s. We run simulations of grains with two materials: amorphous SiO$_2$ and an amorphous \cje{Mg--Fe} silicate of composition Mg$_{1.3}$Fe$_{0.3}$Si$_{1}$O$_{3.6}$, suggested by observations constrained with the ``astrodust'' model of \citet{draine2021dielectric}. With these simulations, we quantify the collision velocity dependence of shattered and vaporized mass fractions, and the resulting size distributions of shattering products.}
   {We find grain shattering thresholds are $\sim$6 km/s for both amorphous SiO$_2$ and astrodust material, which is a factor of $\sim$2 higher than the canonical value for silicates of 2.7 km/s from \citet{Jones1996ApJ...469..740J}. \cje{} We find that the size distributions of shattered products are generally not consistent with the power law distributions predicted by this previous model, \cje{potentially} due to its continuum-dynamical and plane-parallel assumptions \cje{or the increased importance of surface free energy for the smallest grains, but the physical reason for this discrepancy remains uncertain}. We also find that their expression for the fraction of grain mass shocked to a given pressure as a function of collision velocity fails to predict the fraction of shattered or vaporized material observed in our numerical simulations, over-predicting the shattered fraction and under-predicting the vaporized fraction, \cje{except for the most massive grains we simulate, where the predicted vaporized fraction is in decent agreement with the numerical results}. The model of \citet{Hirashita2013EP&S...65.1083H} for the same quantities similarly fails to match the simulations.}
   {We provide updated shattering velocity thresholds for \cje{nanoscale silicate grains} to the astrophysics community. Broadly, our updated threshold velocity prescription suggests that \cje{the smallest} astrophysical dust grains composed of silicate materials may be more robust to shattering in the interstellar medium than previously assumed.}

   \keywords{Cosmic dust --
                Interstellar Medium (ISM) --
                Silicate Nanoparticles --
                Molecular dynamics simulations --
                ReaxFF --
                GFN1-xTB
               }

   \maketitle
%

\section{Introduction}
Dust grains play a central role in the astrochemistry and astrophysics of the interstellar medium \citep[ISM,][]{lazarian2004dust, draine2003interstellar}. These solid particles account for only about 1$\%$ of the total ISM mass, yet are ubiquitous in astrophysical environments \citep{galliano2018interstellar}. The cosmic dust interacts with the gas phase, radiation fields, cosmic rays, and magnetic fields, strongly affecting the composition, thermodynamics, and optical properties of the ISM. They provide catalytic surfaces for molecular hydrogen production and other chemical reactions, and shield these molecules against destruction by UV photons \citep[see][]{Draine2011piim.book.....D}, ultimately allowing interstellar gas clouds to cool and collapse under their own gravity to form stars \citep{GloverClark2012MNRAS.421....9G}.

Cosmic dust is therefore a key component of the galactic ecosystem, but many unknowns remain regarding the basic properties and governing physical processes. This is because empirical constraints on its properties from observations and laboratory experiments face considerable challenges, and predictions of grain physics processes are complicated by the large parameter space of relevant quantities resulting from the many-degree-of-freedom nature of these solid-state clusters. However, a combination of direct and indirect observational evidence, as well as theoretical considerations, have produced a canonical picture in which these grains are primarily composed of refractory, cosmically abundant elements, broadly classifiable as hydrocarbons and silicates \citep{draine2003interstellar}. Their optical properties imply characteristic grain radii of $\sim 0.1\mu$m, with a broad and possibly multi-modal distribution from 0.5~nm to  1~$\mu$m \citep{Hensley2023ApJ...948...55H, Ysard2024A&A...684A..34Y}. More detailed properties of grains are even more uncertain, but best estimates from starlight polarized by grains aligned with interstellar magnetic fields suggest that, at least in the solar neighborhood, grains are only moderately oblate and largely compact \citep[i.e. non-porous,][]{Draine2021ApJ...919...65D, Draine2024ApJ...969...92D}.

Progress toward understanding the processes that determine the cosmic dust grain population requires increasingly realistic physical models. Collisions between two grains are one such process. This can result in non-adhesive bouncing, coagulation (sticking together to form a larger grain), shattering (the breaking of colliding grains into smaller grains), or vaporization (the return of grain mass to the gas phase), as well as changes in the material properties and structure of the grains. We can make a simple order-of-magnitude estimate for the timescale on which an $a_D=0.1~\mu$m size grain will collide with another grain of similar size with $\tau \sim (n_D \sigma_D v_D)^{-1}$ where $n_D$ is the dust number density, $\sigma_D$ is the grain cross-section, and $v_D$ is the dust velocity. Using $n_D \sim D \mu m_H n_g/m_D$ with dust mass fraction $D \approx Z_{\odot}/2$, gas mean molecular weight $\mu \approx 1$ in units of the mass of the hydrogen atom $m_H$,  dust grain mass $m_D \sim 10^{-14}{\rm g}$ (assuming grain material density $\rho_D = 2~{\rm g}~{\rm cm}^{-3}$), and gas number density $n_g$, grain geometric cross section $\sigma_D = \pi a_D^2$, and $v_D\sim 1-10~{\rm km/s}$ \citep{Yan2004-hp}, suggests that grain-grain collisions will regularly occur on astrophysically short timescales: $\lesssim 0.5$ Myr in the dense molecular phase of the ISM (where $n_g\sim10^3~{\rm cm}^{-3}$), $\lesssim 5$ Myr in the diffuse molecular ($n_g\sim10^2~{\rm cm}^{-3}$),  $\lesssim 20$ Myr in the cold neutral media ($n_g\sim30~{\rm cm}^{-3}$), and $\lesssim 800$ Myr in the warm neutral medium \citep[$n_g\sim0.6~{\rm cm}^{-3}$, see][for numerical values]{Draine2011piim.book.....D}. These timescale estimates are upper limits because they neglect the effects of the full dust grain size distribution -- in which most grains are smaller than $0.1~\mu$m -- and because of the local density and velocity enhancements expected in interstellar shocks \citep{McKee1987ApJ...318..674M, Jones1994ApJ...433..797J, Jones1996ApJ...469..740J}. 

These timescales are all shorter than the age of the universe after the Epoch of Reionization \citep{Planck2020A&A...641A...6P}, and for some interstellar phases are comparable to or shorter than the other relevant timescales of interstellar gas and grain evolution such as molecular cloud disruption \citep[10--30 Myr,][]{Chevance2020MNRAS.493.2872C}, stellar dust production through grain nucleation \citep[40 Myr--1 Gyr,][]{SchneiderMaiolino2024A&ARv..32....2S}, grain growth due to gas-phase accretion \citep[5--100 Myr,][]{Hirashita2000PASJ...52..585H, Feldmann2015MNRAS.449.3274F}, grain destruction due to sputtering in the hot ionized medium of the ISM \citep[$\lesssim$400 Myr,][]{Draine2009ASPC..414..453D}, and theoretically estimated grain residence times in the different phases of the ISM \citep[1--40 Myr depending on phase,][]{Peters2017MNRAS.467.4322P}. Each cosmic dust grain in the molecular and neutral phases (i.e. the vast majority of the mass) of the ISM therefore plausibly collides, on average, with a similar mass of other dust grains on the relevant dynamical timescales, and grain-grain collisions likely impact the grain population significantly. Observationally, the optical properties of dust grains -- particularly their thermal emission in the far-infrared and its wavelength dependence -- suggest increasingly larger grains in the denser phases of interstellar gas, especially molecular and protostellar cores \citep{Pagani2010Sci...329.1622P, Dartois2024NatAs...8..359D}, which has been claimed as evidence of grain growth through coagulation \citep{HirashitaLi2013MNRAS.434L..70H}. 

Interstellar grain collision theory therefore plays a central role in modeling dust evolution on astrophysical scales, with \citet{Tielens1994ApJ...431..321T} and \citet{Jones1996ApJ...469..740J} providing widely used treatments. The former paper derived the threshold collision velocities of grains for each possible outcome, as well as the grain mass fraction of each outcome as a function of impact velocity. The latter paper applied this result to candidate materials for interstellar grains and derived the size distribution of grain shattering products. One of their key results is the prediction that grains begin shattering at $\sim$ km/s collision velocities. Specifically, they give values of $2.7$ and $1.2$ km/s for shattering of silicate and carbonaceous grains respectively, and 19 and 23 km/s for vaporization of the same materials. These values have been widely used in models for the evolution of dust in the interstellar medium \citep[e.g.][]{Hirashita2009MNRAS.394.1061H, Asanon2013MNRAS.432..637A, Aoyama2017MNRAS.466..105A, McKinnon2018MNRAS.478.2851M, Huang2021MNRAS.501.1336H, Gjergo2018MNRAS.479.2588G, Li2021MNRAS.507..548L, Parente2022MNRAS.515.2053P, Narayanan2023ApJ...951..100N, Dubois2024A&A...687A.240D}. An alternative formulation from \citet{Hirashita2013EP&S...65.1083H} based on an ansatz for the velocity dependence of shattering presented in \citet{KobayashiTanaka2010Icar..206..735K} fit to match the \citet{Jones1996ApJ...469..740J} results has also been similarly used \citep[e.g.][]{HirashitaAoyama2019MNRAS.482.2555H, HirashitaMurga2020MNRAS.492.3779H, Huang2021MNRAS.501.1336H}.

However, there are many further questions about the physics of dust grain collisions which  \citet{Tielens1994ApJ...431..321T} and \citet{Jones1996ApJ...469..740J} themselves acknowledge are not addressed -- their derivations take a continuum/fluid dynamical approach to estimating grain collision dynamics, the assumptions of which (that grains are much larger than their individual atomic constituents) are less well justified for the smallest grains that likely exist in the ISM (with radii $\sim 0.5 - 5\;{\rm nm}$, sometimes termed ``nano-silicates'' and ``carbon nano-particles''). They also adopt the geometry of a small grain colliding with a plane-parallel much larger grain's surface, which is not valid, in principle, for collisions between grains of similar size. The tools of modern computational chemistry coupled to ever-increasing supercomputing power now enable much more realistic modeling with Molecular Dynamics (MD) simulations that can predict this process atom-by-atom for the relevant grain sizes, which is the focus of the present paper.

A few previous studies have investigated the properties and dynamics of nano-silicates in the context of astrophysical dust with MD \citep{Quadery2017ApJ...844..105Q, Nietiadi2017PCCP...1916555N, Nietiadi2020NRL....15...67N, Nietiadi2022ApJ...925..173N, Alfaridzi2023Icar..39115352A}, but none have been applied to sufficiently high velocities to study \cje{the shattering of compact silicate grains}. Also, relevant laboratory experiments are difficult because of the challenge of mimicking both the relevant interstellar environment and grain dynamics, as well as measuring the properties of such small grains moving so rapidly \citep[e.g.][]{Poppe2000ApJ...533..454P, Whizin2017ApJ...836...94W}. A molecular dynamics approach to this problem is therefore timely and appropriate.

In Section \ref{sec:comp_methods} we describe the computational methods we use to establish model grain structures and perform MD simulations of grain-grain collisions. In Section \ref{sec:results} we present the numerical results of our simulations of grain-grain collisions. We discuss and interpret their implications in Section \ref{sec:discussion}, and conclude in Section \ref{sec:conclusion}.

\begin{figure}
  \centering
  \begin{subfigure}{0.48\linewidth}
    \centering
    \includegraphics[width=\linewidth]{figures/s1.pdf}
    \caption{Si$_{19}$O$_{34}$ (\textbf{S1})}
  \end{subfigure}
  \hfill
  \begin{subfigure}{0.48\linewidth}
    \centering
    \includegraphics[width=\linewidth]{figures/s2.pdf}
    \caption{Si$_{202}$O$_{362}$ (\textbf{S2})}
  \end{subfigure}

  \vspace{0.5em} 

  \begin{subfigure}{0.48\linewidth}
    \centering
    \includegraphics[width=\linewidth]{figures/s3.pdf}
    \caption{Si$_{2839}$O$_{5607}$ (\textbf{S3})}
  \end{subfigure}
  \hfill
  \begin{subfigure}{0.48\linewidth}
    \centering
    \includegraphics[width=\linewidth]{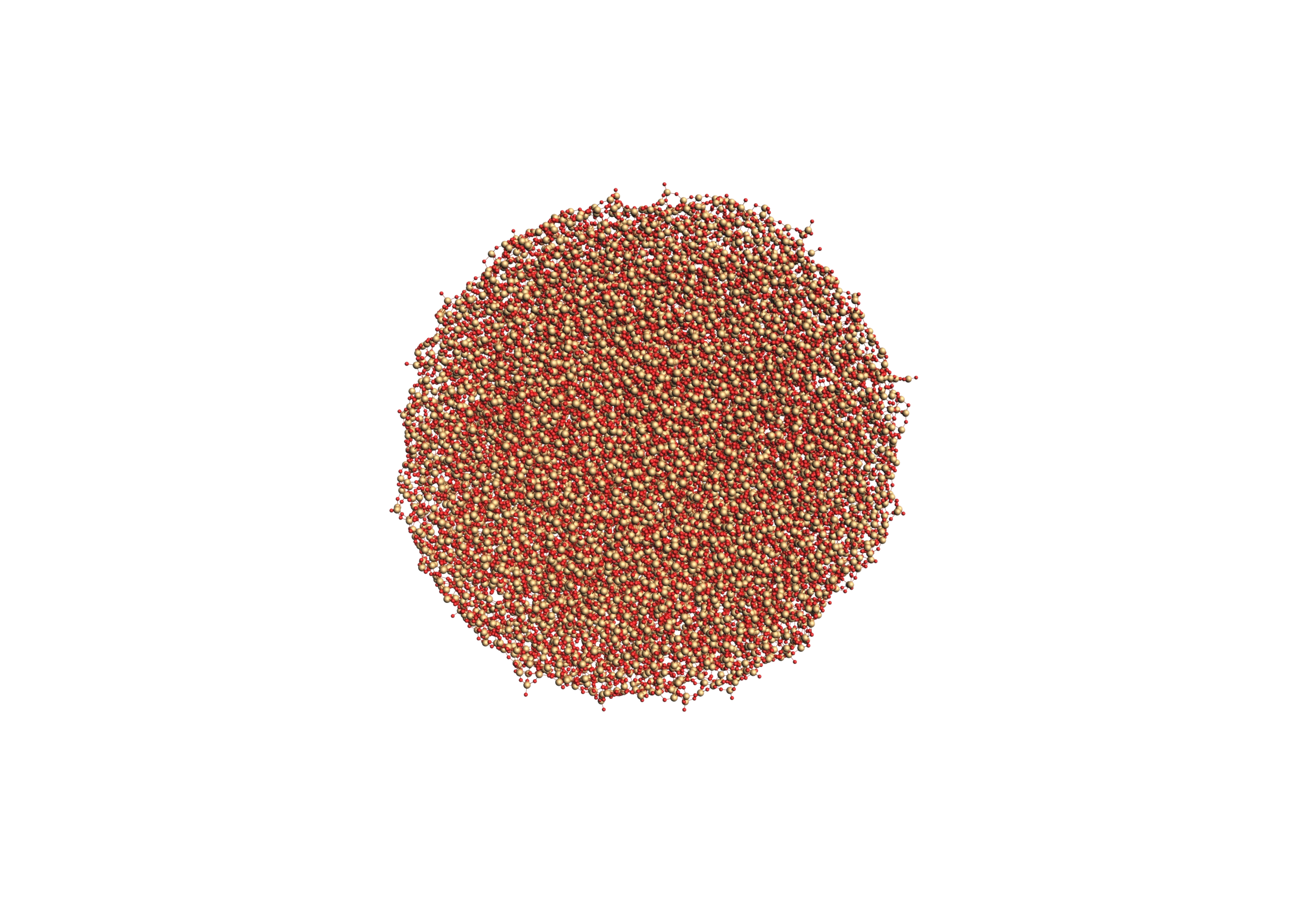}
    \caption{Si$_{8316}$O$_{16664}$ (\textbf{S4})}
  \end{subfigure}
  \caption{Silica SiO$_2$ structure candidates relaxed at 10 K. The structures are obtained using the procedure detailed in Section \ref{subsec:preparation_SiO2_grains}.}
  \label{silica_structures}
\end{figure}

\section{Computational Methods}
\label{sec:comp_methods}

This section describes the preparation of the initial dust grains, simulation details, and the data processing procedure. We employ classical molecular dynamics (MD) simulations in which Newton's equations of motion govern the evolution of the system. Chemical bond formation and breaking are modeled with the reactive force fields (ReaxFF) and the Geometry, Frequency, Noncovalent, extended Tight-Binding (GFN1-xTB) methods. ReaxFF describes bond rearrangements through an empirical bond-order formalism, while GFN1-xTB provides an approximate quantum-mechanical description of the electronic structure. Both ReaxFF and GFN1-xTB methods can capture reactive processes efficiently for large and highly dynamic molecular systems. While these methods generally neglect or only approximately account for nuclear quantum effects such as tunneling and zero-point vibrational energy, as well as nonadiabatic excited-state processes such as electronic excitations -- more accurate methods are not computationally feasible considering the sizes of the molecular systems involved. All calculations are performed by the available engines in the Amsterdam Modeling Suite (AMS) driver program \citep{ams}.

\subsection{Molecular dynamics details}

Grain-grain collision processes involve bond formation and breaking, demanding potentials able to mimic such chemical reactions. Depending on the composition of the dust grains under study, interactions between atoms are modeled employing the ReaxFF potential \citep{van2001reaxff,chenoweth2008reaxff,buehler2006multiparadigm} or the GFN1-xTB semi-empirical quantum mechanical method \citep{grimme2017robust}.

\subsubsection{ReaxFF}
The reactive force field ReaxFF, Eq. \ref{eq:01} is a mathematical model that describes the potential energy of a molecular system by specifying how atoms interact through bonded and non-bonded forces, and is formed of several terms contributing to the potential energy.

\begin{equation}
\begin{split}
E_{\textup{system}} &= E_{\textup{bond}} 
+ E_{\textup{lone \,pair}} 
+ E_{\textup{overcoordination}} 
+ E_{\textup{undercoordination}} \\
&\quad + E_{\textup{valence \,angle}} 
+ E_{\textup{penalty}} 
+ E_{\textup{three \,body \,conjugation}} \\
&\quad + E_{\textup{triple \,bond \,energy}} 
+ E_{\textup{torsional}} 
+ E_{\textup{four \,body \,conjugation}} \\
&\quad + E_{\textup{hydrogen \,bond}} 
+ E_{\textup{vd \,Waals}} 
+ E_{\textup{Coulomb}}.
\end{split}
\label{eq:01}
\end{equation}

The terms in eq. \ref{eq:01} define different parts of the interatomic potential and are thoroughly discussed in the original work by \citet{van2001reaxff} and the Supplementary Information in \citet{chenoweth2008reaxff}. ReaxFF has been extensively applied to investigate bond formation and breaking in a wide range of systems, from combustion chemistry \citep{ashraf2017extension,bertels2020benchmarking,chen2023recent} to astrochemical environments \citep{hashemi2023reaxff,bossion2024accurate,izadi2020reactive,meng2023evolution}. These simulations rely heavily on the availability of a force field that is specifically parameterized for the chemical composition of the molecular system under investigation. In our study, we employ the parameter set \cje{for Si/O materials of} \citet{buehler2006multiparadigm}.

\subsubsection{GFN1-xTB}

GFN1-xTB, developed by \citet{grimme2017robust}, belongs to the extended tight-binding (xTB) family of semiempirical quantum chemical methods. As the name suggests, it solves a simplified tight-binding Hamiltonian to efficiently compute molecular structures, vibrational frequencies, and noncovalent interactions. The total energy in GFN1-xTB is expressed as

\begin{equation}
E_{\text{GFN1-xTB}} = E_{\text{SCC-TB}} + E_{\text{rep}} + E_{\text{dis}} + E_{\text{XB}} + G_{\text{Fermi}},
\label{eq:02}
\end{equation}
where $E_{\rm SCC-TB}$ is the electronic energy, evaluated using an effective self-consistent charge (SCC) tight-binding Hamiltonian \citep{elstner1998self,grimme2017robust}, $E_{\rm rep}$ is the short-range pairwise repulsion term \citep{grimme2017robust}, $E_{\rm dis}$ is the dispersion correction \citep[D3 with Becke–Johnson damping,][]{grimme2011effect}, $E_{\rm XB}$ is the halogen-bond correction term, and $G_{\text{Fermi}}$ is the finite electronic-temperature contribution arising from Fermi smearing.

Due to its much greater computational efficiency relative to the fully quantum-mechanical Density Functional Theory (DFT), GFN1-xTB is well-suited for treating large molecular systems. However, molecular dynamics (MD) simulations using GFN1-xTB remain significantly slower than those based on ReaxFF. As a result, GFN1-xTB-based MD simulations are only possible for our \cje{ADSil} grains, which are explained below.

\subsection{Grain preparation}
\label{subsec:grain}

The focus of the current study is on silicate dust grains, for which we explore two representative chemical compositions: pure amorphous silica SiO$_2$ grains and ``astrodust'' grains with the silicate composition  Mg$_{1.3}$Fe$_{0.3}$Si$_{1}$O$_{3.6}$ inferred from the model of \citet{draine2021dielectric}. We refer to the latter grains with the alias ADSil for the remainder of the paper.

\subsubsection{SiO$_2$ Grains}\label{subsec:preparation_SiO2_grains}

The pure amorphous silica (SiO$_2$) grains are prepared through the following steps: 1) Initial spherical grains of various sizes are extracted from a cubic \cje{silica} bulk in the AMS molecular builder \citep{ams}. 2) ReaxFF MD simulations are employed to minimize the energy of the initial structures at elevated temperatures up to 2000 K. 3) Stepwise energy relaxation down to 10 K results in an amorphous structure which is selected for collision modeling.

The SiO$_2$ dust grain structures generated using this procedure are shown in Figure~\ref{silica_structures}. These grains are constructed to be spherical, non-porous, and amorphous, aiming to realistically represent interstellar medium (ISM) grains \citep{hensley2021observational, draine2024aggregates}. The chemical compositions of the structures are Si$_{19}$O$_{34}$, Si$_{202}$O$_{362}$, Si$_{2839}$O$_{5607}$, and Si$_{8316}$O$_{16664}$ which we label as \textbf{S1}, \textbf{S2}, \textbf{S3}, and \textbf{S4}, respectively. We note that these grains are not stoichiometrically exactly SiO$_2$ (with Si:O ratios of 1:1.789 for \textbf{S1}, 1:1.792 for \textbf{S2}, 1:1.975 for \textbf{S3}, and 1:2.004 for \textbf{S4}), but we nonetheless refer to these grains as being composed of SiO$_2$ throughout the paper. As we show in the following sections, we see no evidence that this variation has an effect on our results, and we note that the uncertainties in real interstellar grain compositions -- as demonstrated by the difference in chemical composition between the two types of grains we simulate -- are much larger than these differences. 

The estimated radii of these grains are approximately 6.5, 16.5, 35, and 50 Å (see Subsection \ref{subsec:data} for details). These grain models therefore only sample the lowest decade in grain sizes expected to exist in the ISM \citep{GuhathakurtaDraine1989ApJ...345..230G, Hensley2023ApJ...948...55H, Ysard2024A&A...684A..34Y}, but this is the limit of computational feasibility for the current study.

\subsubsection{Silicate Astrodust (ADSil) Grains}\label{subsec:preparation_ADSil_grains}

\begin{figure}
  \centering
  \begin{subfigure}{0.48\linewidth}
    \centering
    \includegraphics[width=\linewidth]{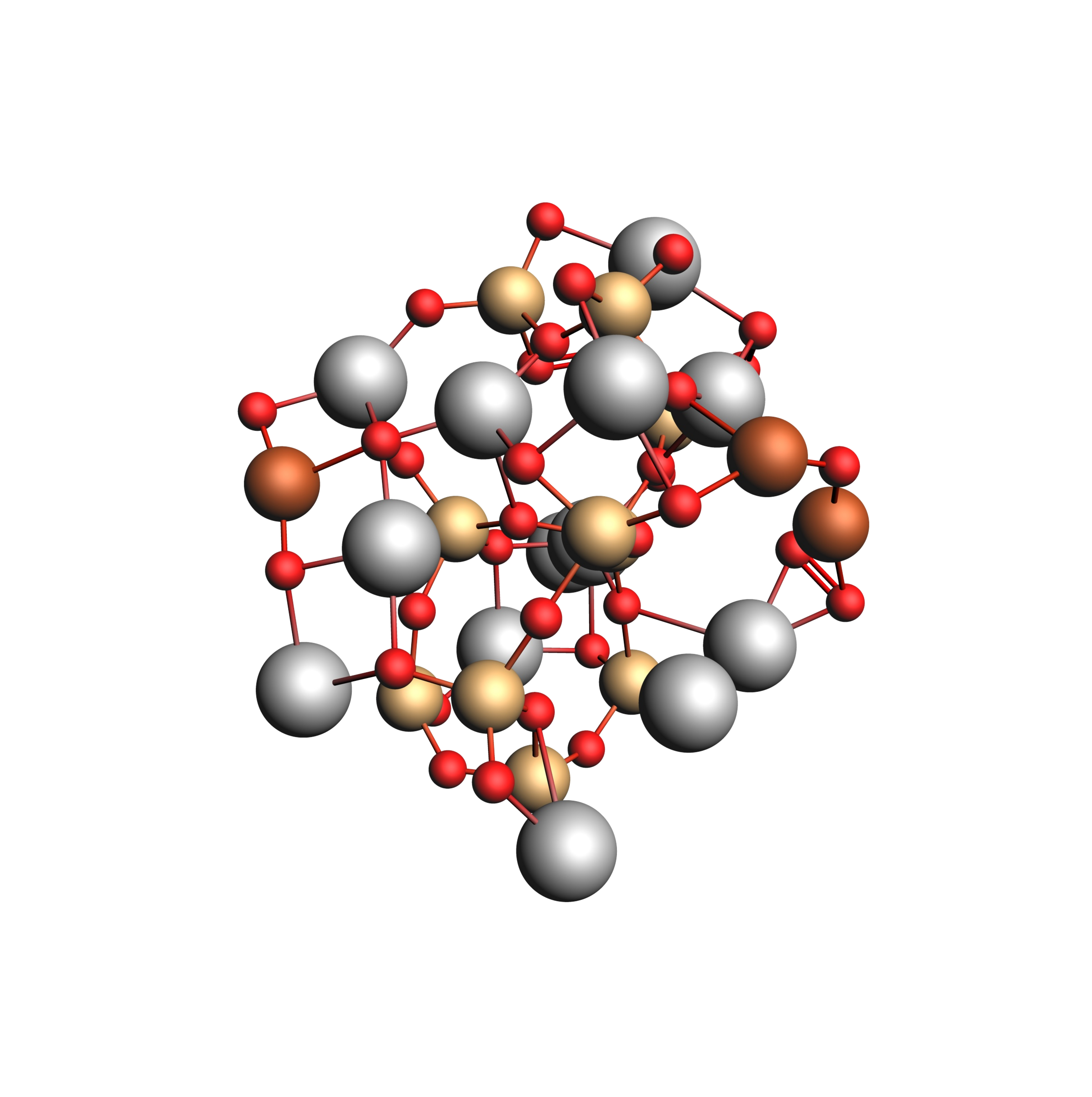}
    \caption{Mg$_{13}$Fe$_3$Si$_{10}$O$_{36}$ (\textbf{A1})}
  \end{subfigure}
  \hfill
  \begin{subfigure}{0.48\linewidth}
    \centering
    \includegraphics[width=\linewidth]{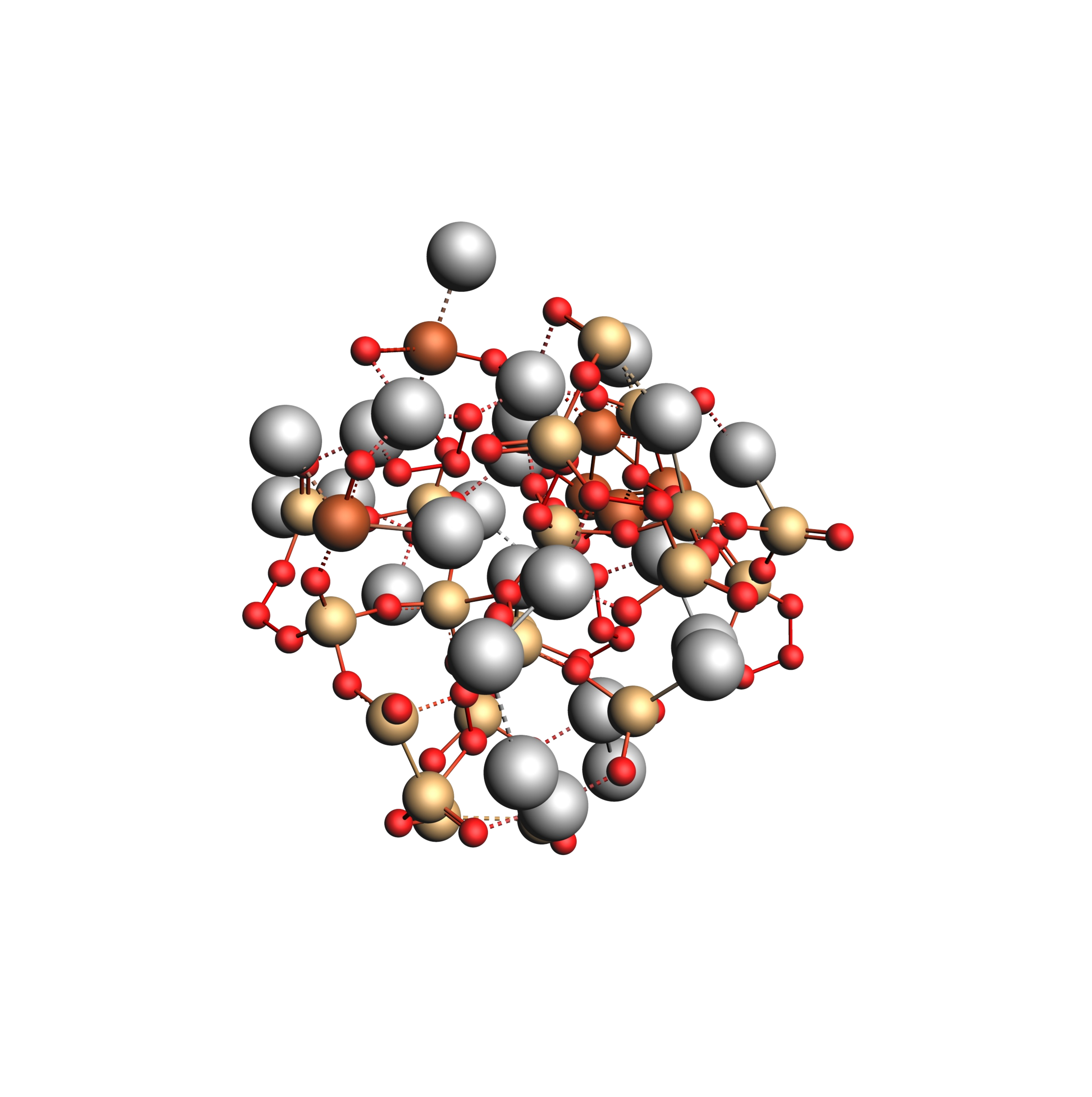}
    \caption{Mg$_{26}$Fe$_6$Si$_{20}$O$_{72}$ (\textbf{A2})}
  \end{subfigure}
  \caption{Astrodust structures optimized using the GFN1-xTB method. The structures are obtained using the procedure  detailed in Section \ref{subsec:preparation_ADSil_grains}.}
  \label{astrodust_structures}
\end{figure}

Silicate astrodust (ADSil) grains were constructed based on the composition proposed by \citet{draine2021dielectric} for \cje{silicate} grain material in the local ISM. To generate representative grain structures, we employed the multi-component artificial force-induced reaction (MC-AFIR) method \citep{sameera2016computational}, as implemented in the Global Reaction Route Mapping GRRM17 code \citep{maeda2013grrm17}. \cje{GFN1-xTB has been successfully applied to a broad range of inorganic systems, including extended inorganic solids and astrochemically relevant clusters, where it provides reliable geometries at a computational cost suitable for large-scale reactive simulations \citep{grimme2017robust, Germain2020, Bannwarth2021, Vicent-Luna2021, Pokora2025, Knattrup2024}. \citet{Komissarov2021} reported deficiencies of the original GFN1-xTB parameterization for organosilicon compounds, in particular for the Si–C/Si–H interactions, but the Si–O and Mg–O bonding that dominates our grains is well described.}

Following the stoichiometry from \citet{draine2021dielectric}, a silicon atom is fixed at the center, while the remaining atoms are randomly distributed around it. \cje{An AFIR parameter (\textgamma) of 100 kJ mol$^{-1}$ was used to generate the artificial force between the central atom and each surrounding atom.} The MC-AFIR search is carried out using the PM6 semiempirical method \citep{stewart2007optimization} for potential energy and derivative calculations performed via the Gaussian16 program \citep{g16}. After collecting 50 structures from the MC-AFIR search, the calculation was stopped, and the resulting structures were collected and reoptimized using the GFN1-xTB method \citep{grimme2017robust} implemented in the AMS DFTB engine \citep{dftb}. The lowest-energy structure from this set is then selected for use in the grain–grain collision simulations.

From this procedure, two ADSil grain structures with chemical formulas Mg$_{13}$Fe$_3$Si$_{10}$O$_{36}$ and Mg$_{26}$Fe$_6$Si$_{20}$O$_{72}$ were generated. These structures, labeled as \textbf{A1} and \textbf{A2}, are shown in Figure~\ref{astrodust_structures}. The estimated radii of \textbf{A1} and \textbf{A2} are 5 and 7.5~\AA, respectively. We do not pursue the investigation of larger astrodust grains in this study due to the lack of available ReaxFF parameters for Mg-containing compounds and the computational expense associated with GFN1-xTB-based molecular dynamics simulations.

\subsection{Grain-Grain Collisions and Data Processing}
\label{subsec:data}
The Molecule Gun feature in the AMS driver program \citep{ams} is used to introduce a dust grain (projectile) into the simulation box, initiating a collision with a preexisting grain (target). The collision trajectory is aligned along the axis connecting the atoms nearest to the centers of the two grains. The initial orientation of the projectile grain is random.  We sample grain relative, i.e. lab frame collision velocities in the range of $v_l = $0.1--20 km/s. We note that these velocities, being in the lab frame of the target dust grain at rest, are different from the center-of mass reference frame of the \citet{Tielens1994ApJ...431..321T} calculations to which we are comparing. This is because their target grain is effectively infinite in mass, so in their set-up the maximum kinetic energy possibly transferred to the grains is the initial projectile grain kinetic energy $\frac{1}{2}M_Pv_r^2$ where $M_P$ is the projectile grain mass and $v_r$ is its relative velocity (which is equivalent in their lab and center-of-mass frames because of the infinite mass of their target grain), while in our set up the maximum kinetic energy that can do work on the grains during coagulation is $\frac{1}{2}M_P\frac{M_T}{M_P + M_T}v_l^2$ where $M_T$ is the target grain mass and $v_l$ is the lab-frame relative velocity. Consequently, a fair comparison to the \citet{Tielens1994ApJ...431..321T} theory requires we scale our velocities by a factor dependent on the grain mass ratio

\begin{equation}
    v_r = \left[\frac{(M_T/M_P)}{1 + M_T/M_P}\right]^{1/2}v_l
\end{equation}

\noindent which we do in all subsequent analysis.

We do not sample multiple orientations for a given velocity and grain pairing because the grains are approximately spherical. This may introduce some scatter into our results but should not bias them because the orientation is random. The initial temperature of each grain is pre-set to be 10 K, but there is no control over the temperature after the simulation starts. A velocity Verlet integrator with a constant time step of 0.25 fs is used. The simulations are performed with $2\times 10^5$ timesteps. Collisions are referred to in the form \textbf{Xi}--\textbf{Xj}, where \textbf{X} represents either \textbf{S} for SiO$_2$ grains or \textbf{A} for  ADSil grains.

This numerical set-up only simulates head-on collisions with impact parameter $b = 0$. This makes the interpretation of our results simpler, because all translational energy is used in the collision to vibrationally excite, break, or make chemical bonds in the grains. We therefore get a direct dependence of collision outcome on collision velocity, which is the goal of this study. However, we acknowledge that in the real ISM many collisions will likely be grazing encounters with non-zero impact parameters. While the effects of this complication are beyond the scope of the current study, we note here that this represents an interesting and important parameter space for exploration in future work.

Data are collected at the end of each simulation by analyzing the final atomic positions. This post-simulation analysis provides the mass distribution, size distribution, and chemical composition of the resulting grains and molecules. Grain sizes are reported by their radii (in \AA). We utilize two definitions for radii. The first is termed ``geometric'' $a_{\rm geo}$, and is estimated based on the spatial distribution of the atoms. For multi-atom clusters or molecules, the diameter is defined as the maximum distance between any two surface atoms (those that lie on the convex hull of the  grain), plus the van der Waals (vdW) radii of the two atoms involved. This captures both the shape and the physical extent of the molecular structure more accurately than point-to-point distances alone. In the case of single atoms, the diameter is approximated as twice the vdW radius, treating the atom as a sphere. This refined geometric approach provides a consistent and physically meaningful measure of grain size across a wide range of chemical compositions and structural configurations. The second is termed ``effective'' $a_{\rm eff}$ and is defined uniquely by the grain mass as the radius of a compact, uniform density sphere

\begin{equation}
    a_{\rm eff} \equiv \left(\frac{3 m}{4\pi\rho}\right)^{1/3}.
\end{equation}

\noindent where $m$ is the grain mass and $\rho$ is the assumed grain density, which we take to be  $1.57\;{\rm g}\;{\rm cm}^{-3}$ for the amorphous SiO$_2$ and $3.10\;{\rm g}\;{\rm cm}^{-3}$ for ADSil, as these are the mass-weighted average densities of our initial model grains for each material. For the SiO$_2$ grains, this is lower than the bulk amorphous silica density of $\sim 2.2\;{\rm g}\;{\rm cm}^{-3}$ \footnote{\url{https://pubchem.ncbi.nlm.nih.gov/compound/Silica\#section=Density}}. We attribute the difference primarily to our geometric radius definition, which includes the van der Waals shell of the outermost atoms and therefore inflates the volume used to compute the density. This effect is largest for our smallest grains, where the vdW shell makes up a substantial fraction of the inferred radius.

Grain radii are used to delineate grains from gas-phase molecules. For the grain-gas threshold we adopt $a = 4.5\mathring{A}$, since smaller grains are expected to be unstable to sublimation by the interstellar radiation field and therefore eventually break apart into their constituent atoms in astrophysical environments \citep{GuhathakurtaDraine1989ApJ...345..230G}. By default we use $a_{\rm eff}$ for this categorization. For the above-mentioned densities this radius corresponds to grain masses of 361.26 amu for SiO$_2$ and 711.69 amu for ADSil. We have checked that our results do not depend sensitively on (i.e. are not qualitatively changed by a factor of 2 increase in) these minimum grain masses.

We note that throughout, we use the term ``disrupt'' to indicate the collisional fragmentation of grains into products of any size, while we use ``shatter'' to refer specifically to the fraction of products that are large enough to be considered grains,  and we use ``vaporize'' to refer to the fraction whose constituents are smaller than our minimum grain size and are therefore returned to the gas phase.

\section{Results}
\label{sec:results}

Here we analyze the outcomes of our simulations, which are the following: \textbf{S1}--\textbf{S1}, \textbf{S2}--\textbf{S2}, \textbf{S3}--\textbf{S3}, \textbf{S4}--\textbf{S4} (``homogeneous'' SiO$_2$), \textbf{A1}--\textbf{A1}, and \textbf{A2}--\textbf{A2} (``homogeneous'' ADSil), \textbf{S1}--\textbf{S3}, \textbf{S2}--\textbf{S3} (``heterogeneous'' SiO$_2$), \textbf{A1}--\textbf{A2} (``heterogeneous'' ADSil), and \textbf{A1}--\textbf{S1} (mixed composition). 

\subsection{Shattering and Vaporization as a Function of Impact Velocity}
\label{subsec:results_shat_vap_velocity}

\begin{figure*}
    \centering
    \includegraphics[width=\linewidth]{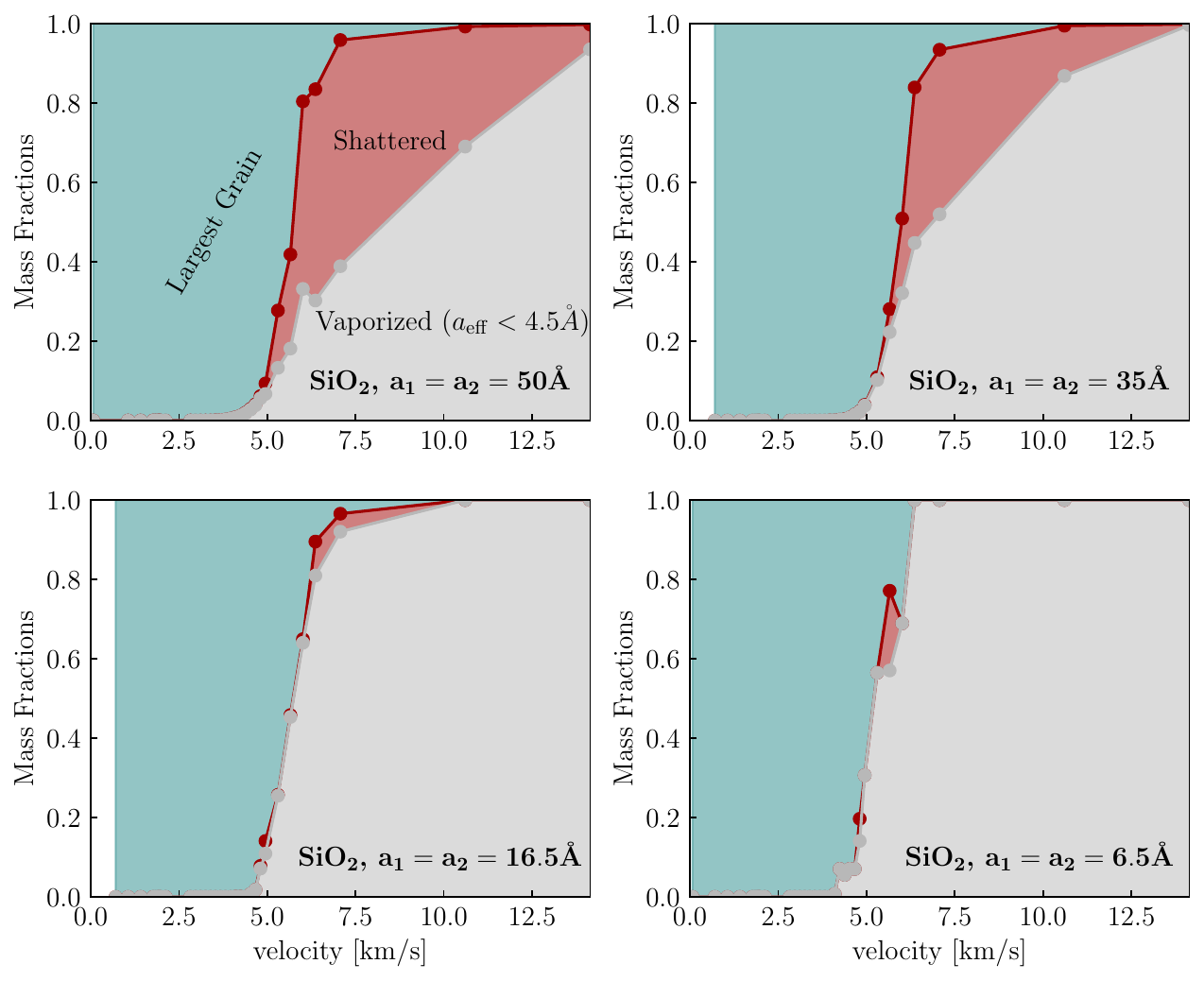}
    \caption{Collision outcomes for SiO$_2$ grains of the same size as a function of impact velocity. Each panel shows the mass fractions, as a fraction of the total combined mass of initial particles, of the 3 main collision outcomes: the largest remaining grain (defined by mass, blue), shattered products smaller than the largest grain but larger than the vaporization limit (red), and the vaporized fraction of collision products smaller than our minimum grain size of 4.5\AA~(gray). Initial grain sizes are indicated by $a_1$ and $a_2$. The top left panel shows results for the \textbf{S4}--\textbf{S4} simulations, top right for \textbf{S3}--\textbf{S3}, bottom left for \textbf{S2}--\textbf{S2}, and bottom right for \textbf{S1}--\textbf{S1}. White regions indicate velocities outside the range simulated.}
    \label{fig:SiO2_homo_f_shat_vap}
\end{figure*}

\begin{figure}
    \centering
    \includegraphics[width=\linewidth]{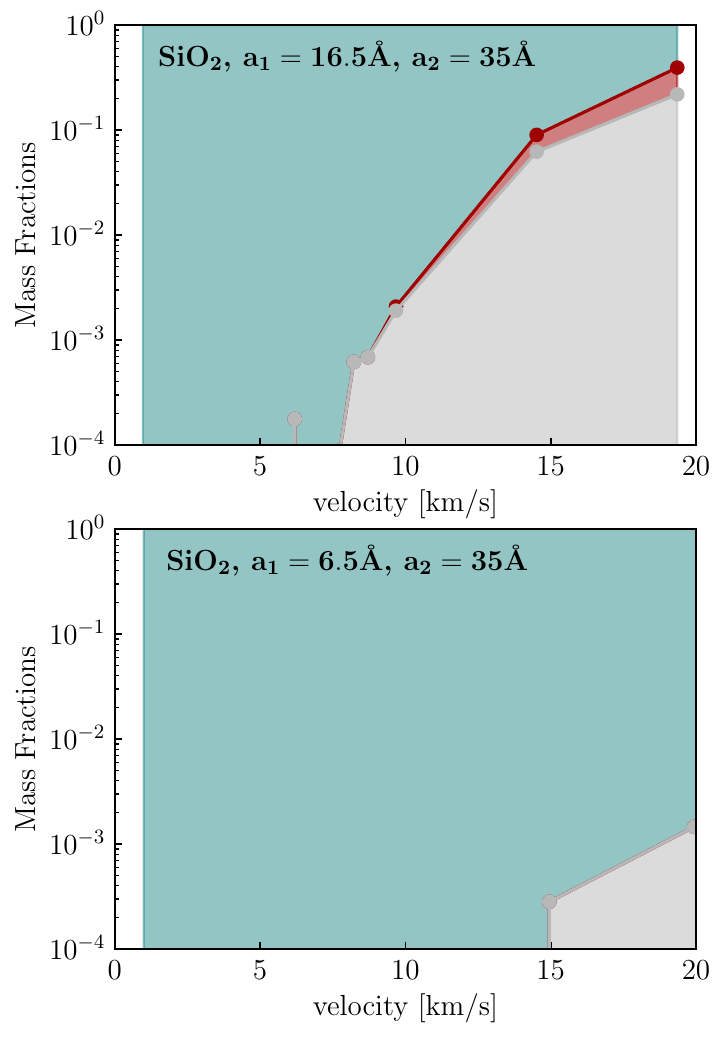}
    \caption{Same as Figure~\ref{fig:SiO2_homo_f_shat_vap} but for SiO$_2$ grains with different sizes. The top panel shows results for the \textbf{S2}--\textbf{S3} simulations, the bottom for  \textbf{S1}--\textbf{S3}.}
    \label{fig:SiO2_hetero_f_shat_vap}
\end{figure}

\begin{figure}
    \centering
    \includegraphics[width=\linewidth]{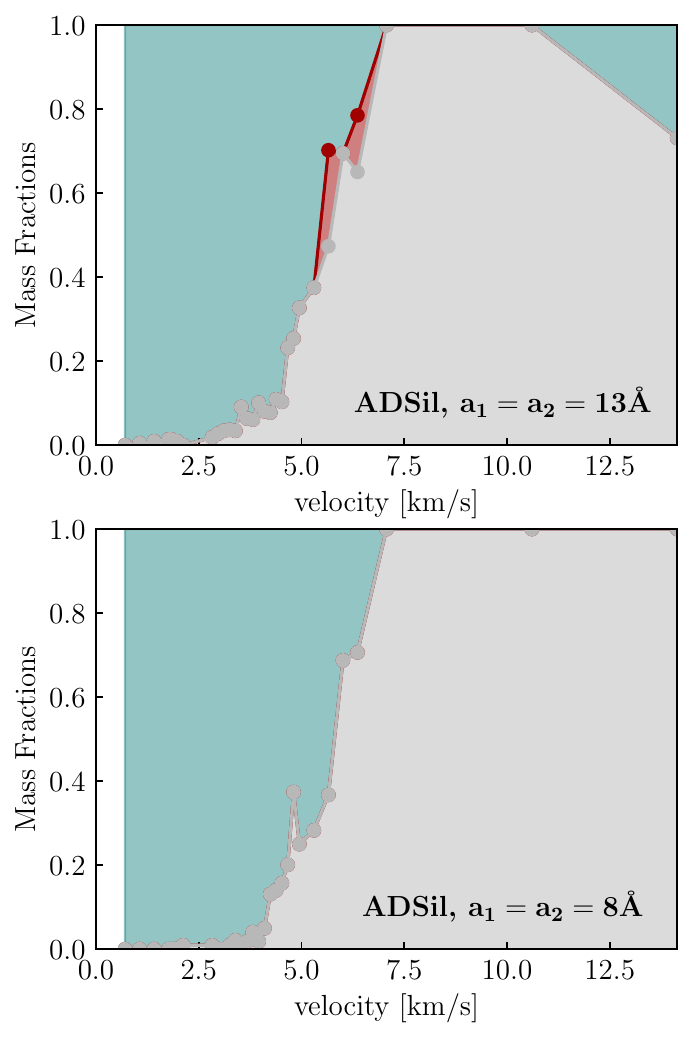}
    \caption{Same as Figure~\ref{fig:SiO2_homo_f_shat_vap} but for astrodust silicate (ADSil) grains of the same size. The top panel shows results for the \textbf{A2}--\textbf{A2} simulations, the bottom for \textbf{A1}--\textbf{A1}.}
    \label{fig:AD_homo_f_shat_vap}
\end{figure}

\begin{figure}
    \centering
    \includegraphics[width=\linewidth]{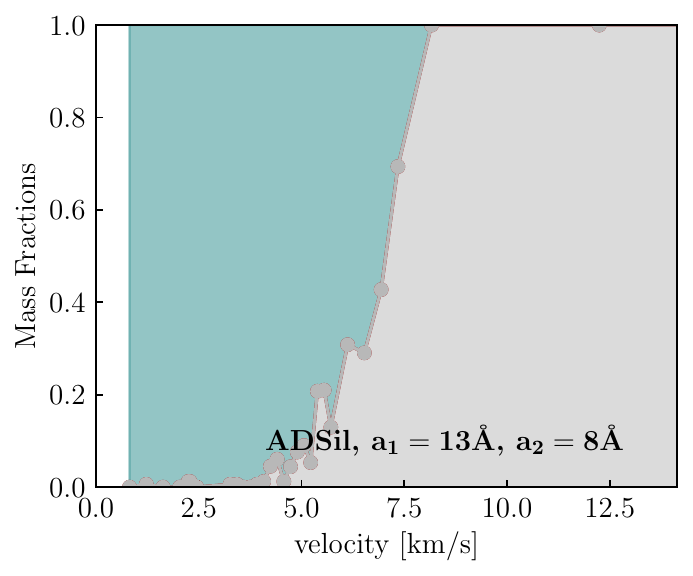}
    \caption{Same as Figure~\ref{fig:SiO2_homo_f_shat_vap} but for ADSil grains with different sizes: \textbf{A2}--\textbf{A1}.}
    \label{fig:AD_hetero_f_shat_vap}
\end{figure}

\begin{figure}
    \centering
    \includegraphics[width=\linewidth]{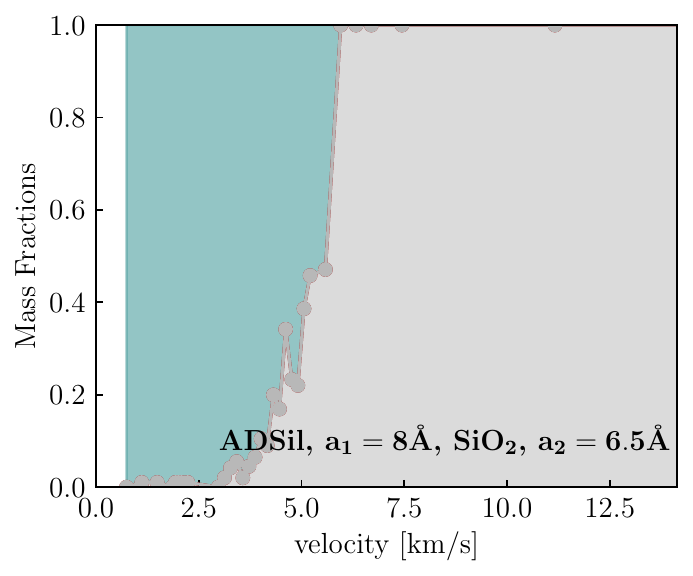}
    \caption{Same as Figure~\ref{fig:SiO2_homo_f_shat_vap} but for mixed composition pairs: \textbf{A1}--\textbf{S1}.}
    \label{fig:mixed_f_shat_vap}
\end{figure}


\begin{table}
\caption{Threshold velocities (km/s). Velocity subscripts indicate the total fraction of combined collision mass used to define that threshold: $v_{0.5}$ indicates the collision velocity at which half of the combined initial grain mass is disrupted or vaporized. Fractions are determined by linearly interpolating between simulation data points. \label{table:threshold_velocities}}
\centering
\begin{tabular}{cccc}
\hline
\hline
Collision \& Outcome & $v_{\rm 0.1}$ & $v_{\rm 0.5}$ & $v_{\rm 0.9}$ \\
\hline
\textbf{S4}--\textbf{S4}, disrupted & 5.0 & 5.7 & 6.7 \\
\textbf{S4}--\textbf{S4}, vaporized & 5.1 & 8.4 & 13.6 \\
\textbf{S3}--\textbf{S3}, disrupted & 5.3 & 6.0 & 6.8\\
\textbf{S3}--\textbf{S3}, vaporized & 5.3 & 6.9 & 11.5 \\
\textbf{S2}--\textbf{S2}, vaporized & 4.9 & 5.7 & 6.9 \\
\textbf{S1}--\textbf{S1}, vaporized & 4.7 & 5.2 & 6.3 \\
\textbf{S2}--\textbf{S3}, disrupted & 14.7 & >19.4 & >19.4\\
\textbf{S2}--\textbf{S3}, vaporized & 15.7 & >19.4 & >19.4 \\
\textbf{A2}--\textbf{A2}, vaporized & 4.0 & 5.7 & 6.9 \\ 
\textbf{A1}--\textbf{A1}, vaporized & 4.2 & 5.8 & 6.8 \\
\textbf{A2}--\textbf{A1}, vaporized & 5.3 & 7.1 & 7.9 \\ 
\textbf{A1}--\textbf{S1}, vaporized & 4.0 & 5.6 & 5.9 \\
\hline
\end{tabular}
\end{table}

We first examine the outcome of our grain-grain collision simulations as a function of impact velocity. Figure \ref{fig:SiO2_homo_f_shat_vap} shows the mass fractions of 3 separate components of the collision products for collisions between same-size SiO$_2$ clusters: the largest grain, the shattered fraction of grains smaller than the largest grain but larger than the threshold size to be considered a grain, and the vaporized fraction of products with radii below the threshold of minimum grain size i.e. the fraction returned to the gas phase. Figure \ref{fig:SiO2_hetero_f_shat_vap} shows results for heterogeneous SiO$_2$ simulations, Figure \ref{fig:AD_homo_f_shat_vap} shows the same for homogeneous collisions between our ADSil grain models, Figure \ref{fig:AD_hetero_f_shat_vap} for heterogeneous ADSil collisions, and Figure \ref{fig:mixed_f_shat_vap} for our mixed-composition simulations. Threshold velocities for given collision outcomes and grain types, determined by various mass fraction definitions, are shown in Table \ref{table:threshold_velocities}. 

Broadly, Figure \ref{fig:SiO2_homo_f_shat_vap} shows that SiO$_2$ grains coagulate for all collision velocities lower than $\approx 5$ km/s -- we do not observe any ``bouncing'' behavior in our simulations. There is then a rapid transition between $5$ and $7$ km/s in which the combined shattered and vaporized fractions increase to approximately 90\%. For increasingly higher velocities, most of the colliding grains are disrupted. For the two largest grain sizes, the fraction that is vaporized increases approximately linearly from $\approx 7$ km/s. In the case of the two smallest grain sizes, which are near the gas phase threshold to begin with, most of the mass of the grains which is disrupted is vaporized. Disrupted fractions for collisions between SiO$_2$ grains of different sizes shown in Figure \ref{fig:SiO2_hetero_f_shat_vap} are predictably less, but show similar velocity dependence. Disruption begins above a similar threshold velocity for the \textbf{S2}--\textbf{S3} simulations, but only begins at much higher values for \textbf{S1}--\textbf{S3}, and very little is disrupted. This demonstrates that large mass-ratio collisions (approximately 100:1 in this case) do not disrupt the larger grain for velocities relevant to grain collisions in the ISM, at least for the small grains we simulate in this study.

For ADSil (Figures \ref{fig:AD_homo_f_shat_vap} and \ref{fig:AD_hetero_f_shat_vap}), some of the same broad trends appear to hold: disruption begins around a slightly lower threshold velocity of $\approx 4$ km/s, and increases rapidly around 5 to 6 km/s.  For these smaller grains, anything not contained in the largest grain is vaporized into the gas phase, which is the entire combined mass for velocities 7 km/s and higher. The same general trends are observed for the mixed composition collision, shown in Figure \ref{fig:mixed_f_shat_vap}. 

In all cases, any reasonable choice for a threshold shattering velocity for either composition and any grain size would be greater than the canonical value of 2.7 km/s from \citet{Jones1996ApJ...469..740J} by a factor of approximately 2. \cje{The source of this discrepancy is still unclear, which we discuss further in Section \ref{subsec:discussion_vel_thresh}.}

\begin{figure*}
    \centering
    \includegraphics[width=\linewidth]{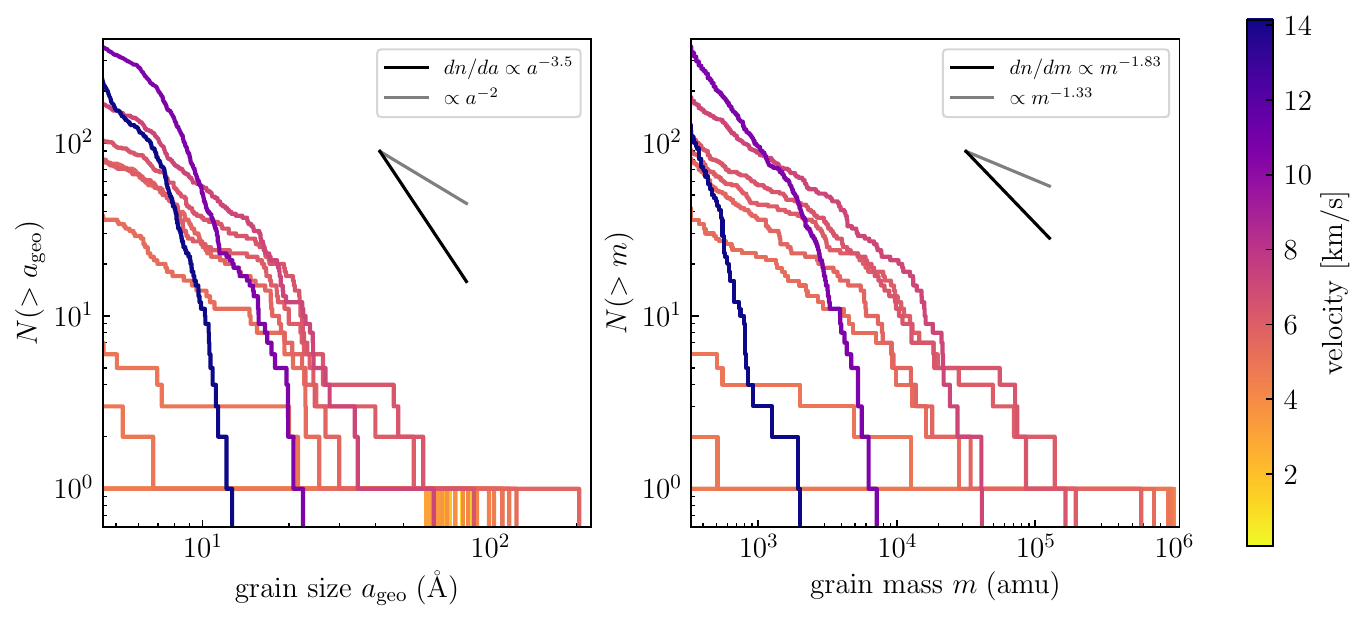}
    \caption{Size distributions of collision products for all simulated collision velocities for the largest SiO$_2$ grains (\textbf{S4}--\textbf{S4}). Reference scalings are shown for size distributions with scalings $dn/da \propto a^{-3.5}$ \citep{MRN1977ApJ...217..425M} and $\propto a^{-2}$. The equivalent scalings are shown for the mass distribution, assuming spherical grains and identical density, in which case $dn/da \propto a^{-\alpha} \implies dn/dm \propto m^{-(\alpha + 2)/3}$ so $dn/da \propto a^{-3.5} \implies dn/dm \propto m^{-1.83}$ and $dn/da \propto a^{-2} \implies dn/dm \propto m^{-1.33}$. Note that these are the exact cumulative distribution functions (CDFs) for the data -- no binning has been applied.}
    \label{fig:s4s4_size_distributions}
\end{figure*}

\subsection{Shattering Product Size Distributions}

Shattering is expected to be a key regulator of the dust grain size distribution \citep[e.g.][]{Hirashita2013EP&S...65.1083H, Li2021MNRAS.507..548L, Huang2021MNRAS.501.1336H, Narayanan_2026}. It is therefore interesting to analyze the size distribution of shattering products in our simulations. These are shown in Figure \ref{fig:s4s4_size_distributions} for the same-size collisions of the largest SiO$_2$ grains i.e. the \textbf{S4}--\textbf{S4} simulations. The inset lines include a reference to $dn/da \propto a^{-3.5}$, the Mathis, Rumpl, \& Nordsieck (MRN)  distribution suggested by observations \citep{MRN1977ApJ...217..425M} -- slightly steeper than the dependence of  $dn/da \propto a^{-3.3}$ predicted by \citet{Jones1996ApJ...469..740J} -- and  $dn/da \propto a^{-2}$ as an additional reference. 

The results are complicated but we can identify several broad trends. With the possible exception of the grain mass distribution at  14 km/s collision velocity, the distributions generally do not appear to be well described by a single power law, and display a complicated dependence on collision velocity. In general the distributions appear to become steeper with increased collision velocity, on average shallower than a power-law exponent $\alpha = 3.5$ (where $\alpha$ is defined by $dn/da \propto a^{-\alpha}$) for velocities up to $\sim$7 km/s but steeper for higher velocities. As well, at the two highest velocities, the largest grain collisions display a multi-modal behavior in the size distribution around the grain-gas limit ($\sim 5$~\AA) that is not present in the mass distribution, suggesting changes in average shape or density with grain size.

\subsection{The Shapes of Collision Products}

\begin{figure}
    \includegraphics[width=\linewidth]{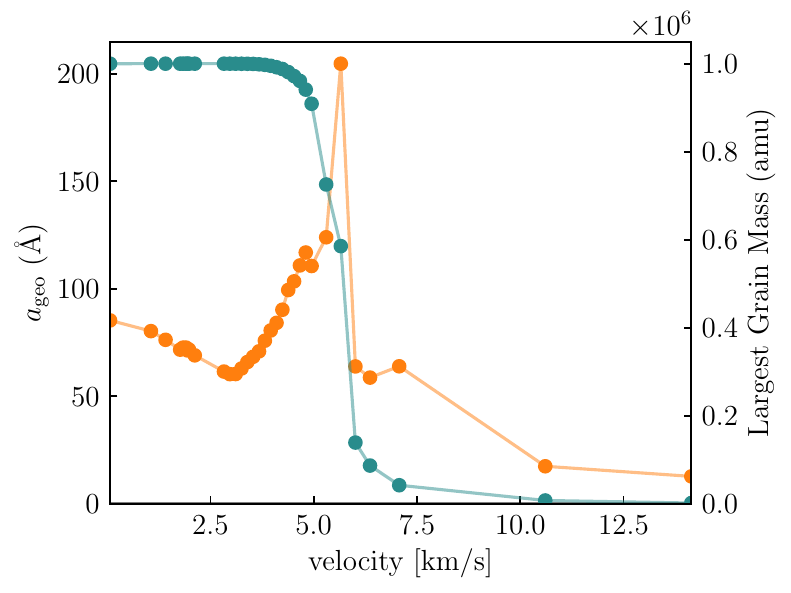}
    \caption{The geometric radius $a_{\rm geo}$ (orange) and mass (teal) of the largest remaining grain in each \textbf{S4}--\textbf{S4} simulation as a function of collision velocity}
    \label{fig:largest_grain_sizes}
\end{figure}

\begin{figure}[htbp]
    \centering  
    \begin{subfigure}{\linewidth}       
        \centering 
        \includegraphics[width=0.5\hsize, angle=90]{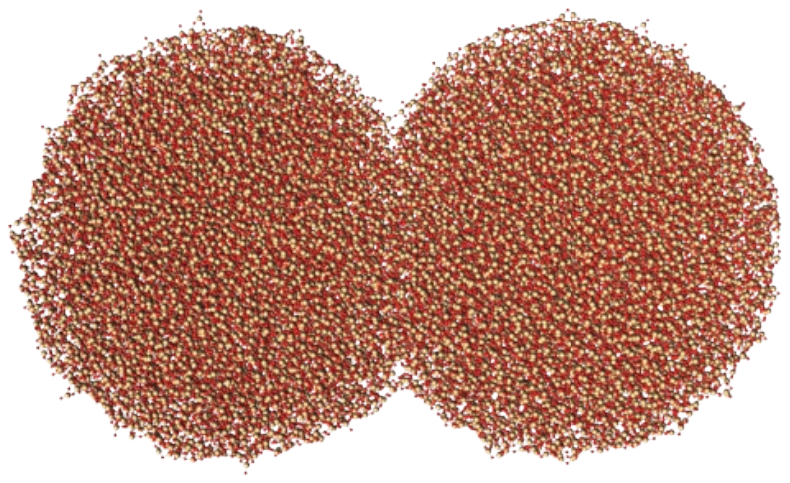}         
        \caption{}
        \label{fig:s4s4_image_low}
    \end{subfigure}
    \begin{subfigure}{\linewidth}
        \centering
        \includegraphics[width=0.5\hsize, angle=90]{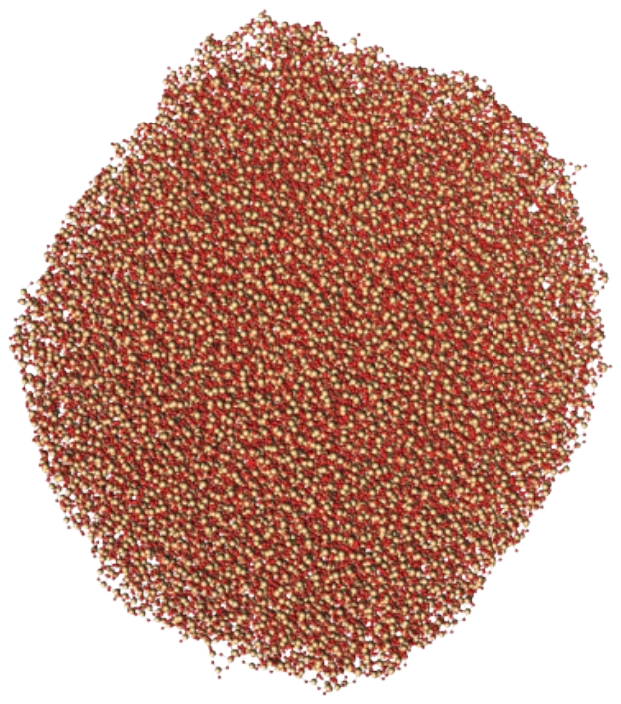}
        \caption{}
        \label{fig:s4s4_image_medium}
    \end{subfigure}        
    \begin{subfigure}{\linewidth}
        \centering
        \includegraphics[width=0.5\hsize, angle=90]{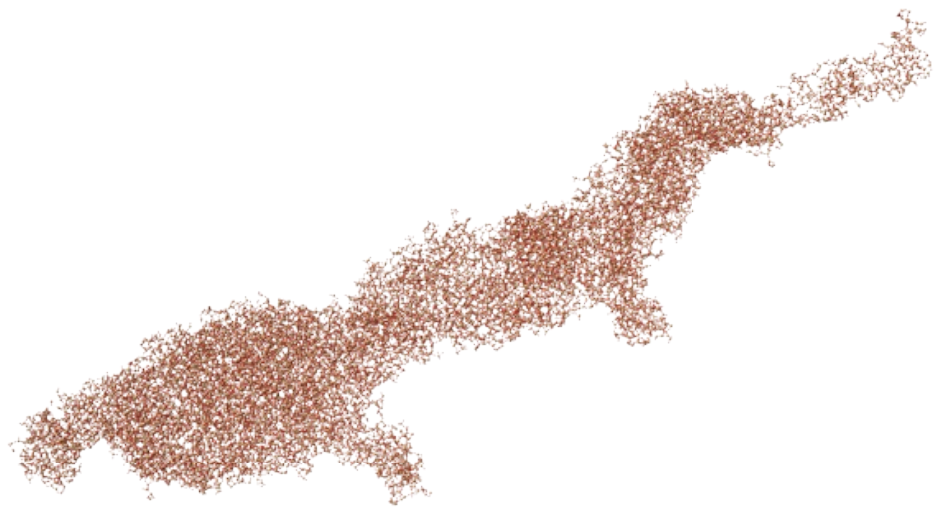}
        \caption{}
        \label{fig:s4s4_image_high}
    \end{subfigure}
\caption{Images of \textbf{S4}--\textbf{S4} coagulation at collision velocities of (a) 1.4 km/s, (b) 3.1 km/s, and (c) 5.7 km/s.}
\label{fig:s4s4_images}
\end{figure}

The shapes of grains -- i.e. their deviations from sphericity -- also affect their optical and dynamical properties. We can gain insight into the effect of grain collisions on their shapes by analyzing the properties of the largest grain: Figure \ref{fig:largest_grain_sizes} shows the size $a_{\rm geo}$ and mass of the largest remaining grain as a function of collision velocity for the \textbf{S4}--\textbf{S4} simulations. At low velocities, the resulting coagulated grain has a geometrically defined size of $a_{\rm geo} \approx 90$~\AA, which is slightly less than the diameter of each colliding grain, suggesting that the grain is composed of two spheres that largely retain their shape but stick together with some overlap, i.e. a prolate grain. Higher collision velocities produce smaller grains, suggesting the constituent grains stick together with larger overlap. This trend reverses at collision velocities of $\sim 3$ km/s, at which point the resulting grain becomes increasingly larger up to the shattering limit at $\sim 5.5$ km/s. This suggests that higher collision velocities increasingly reshape the coagulated grain, resulting in an elongated, increasingly prolate geometry, until the shattering limit at which point the largest resulting grain rapidly becomes smaller with higher impact velocity.  This phenomenology is demonstrated in images of the largest remaining grain for different velocities from the \textbf{S4}--\textbf{S4} simulations shown in Figure \ref{fig:s4s4_images}.

\section{Discussion}\label{sec:discussion}

While other studies have presented the analysis of silicate nanoparticle models with modern computational chemistry techniques \citep[e.g.][]{VanHoang2007}, even with the similar motivation of studying astrophysical dust \citep[e.g.][]{MarinosoGuiu2026}, to our knowledge there are no results in the published literature on what happens when they collide at km/s speeds studied with molecular dynamics simulations. The only existing calculations to which we can compare our results are the empirically informed symbolic calculations of \citet{Tielens1994ApJ...431..321T} and \citet{Jones1996ApJ...469..740J} in the astrophysical dust context, and \citet{KobayashiTanaka2010Icar..206..735K} in the context of asteroids and planetesimals extended by \citet{Hirashita2013EP&S...65.1083H} to model interstellar dust grains. We confront their predictions with our numerical results.

\subsection{Shattering and Vaporization Velocity Thresholds}\label{subsec:discussion_vel_thresh}
In \citet{Tielens1994ApJ...431..321T} and \citet{Jones1996ApJ...469..740J}, a grain collision speed shattering threshold of 2.7 km/s is quoted for silicate materials, a factor of $\approx 2$ smaller than found in our simulations, regardless of possible definitional ambiguities (see Table \ref{table:threshold_velocities} and Section \ref{subsec:results_shat_vap_velocity}). They calculate this velocity by equating the initial pressure to which the target grain material is shocked (derived from continuum, fluid dynamical considerations) to the shear modulus of the grain material, \cje{resulting in the following expression:}

\begin{equation}\label{eq:Tielens1994_2_16}
   \cje{\frac{P_{1i}}{\rho_0 c_0^2} = \left(s + \frac{1 + \mathscr{R}}{\mathscr{M}_r}\right)\frac{\mathscr{M}_r^2}{(1 + \mathscr{R})^2}}
\end{equation}

\noindent \cje{where $P_{1i}$ is the target grain initial shock pressure, $\rho_0$ is the target material density, $c_0$ is the target material sound speed, $\mathscr{M}_r \equiv v_r/c_0$ is the impact velocity $v_r$ Mach number in the target, $\mathscr{R} \equiv \sqrt{\frac{s\rho_0}{s_P \rho_{0P}}}$ is a function of the target and projectile (subscript $P$) material properties where $s$ is defined by an empirically calibrated relationship between the shock velocity and the velocity of the shocked matter $v_s = c_0 + s v_1$. For the shattering threshold, \citet{Jones1996ApJ...469..740J} takes the shear modulus of the material, $G = 3\times 10^{11}$ dyn cm$^{-2}$ for silicates, as the critical pressure. \citet{Tielens1994ApJ...431..321T} starts with the {\it velocity} threshold for vaporization as $v_t = 4 \epsilon_b^{1/2}$ where $\epsilon_b$ is the binding energy of the material (per mass), taken to be 2.9 $\times 10^{11}$ ergs g$^{-1}$ for silicates, from which the vaporization pressure of $P_v = 5.4\times 10^{12}\;{\rm dyn}\;{\rm cm}^{-2}$ is obtained from eq. \ref{eq:Tielens1994_2_16}, eq. 2.16 in \citet{Tielens1994ApJ...431..321T}. Applying this threshold pressure to our materials far over-predicts the vaporization thresholds \cje{we measure in the simulations}, which evaluate to 26.7 km/s for SiO$_2$ and 34.2 km/s for astrodust.}

The disagreement between their framework and our results suggests the magnitude of other expected effects, such as the increased relative importance of surface energy and the reduced number of degrees-of-freedom in vibrational modes for small grains. Further molecular dynamical simulations with both larger grains and grains of different compositions could help to \cje{clarify this discrepancy}. 

Finally, we note that \citet{Shull1977ApJ...215..805S} estimated a disruption threshold for colliding grains with

\begin{equation}\label{eq:Shull1977}
    v_{\rm thr} = \sqrt{\frac{2E_{\rm bind}}{\bar{m}}}
\end{equation} 

\noindent where $E_{\rm bind}$ is the binding energy per molecule and $\bar{m}$ is the mean molecular weight of the material. Using recent estimates of silicate grain binding energies from \citet{Hansson2026A&A...707A..54H} in this equation gives $v_{\rm thr} = 7.9$ km/s for an FeMgSiO$_4$ composition \citep[as was assumed in][]{Hansson2026A&A...707A..54H}, $v_{\rm thr} = 10.0$ km/s for SiO$_2$ and $v_{\rm thr} = 8.4$ km/s for the ADSil compositions. We note that the latter two of these estimates are not self-consistent because they use binding energies calculated for a different material, but absent more accurate binding energy calculations we use these as a best-guess. Nonetheless, with these binding energies, the \citet{Shull1977ApJ...215..805S} estimate does fairly well when compared to our molecular dynamics simulations, especially when Eq. \ref{eq:Shull1977} is interpreted as a vaporization threshold for grains with size $a \gtrsim 30$~\AA~(i.e. our \textbf{S4}--\textbf{S4} and \textbf{S3}--\textbf{S3} simulations, see Figure \ref{fig:SiO2_homo_f_shat_vap} and Table \ref{table:threshold_velocities}). This is the most physically appropriate interpretation of Eq. \ref{eq:Shull1977}, because the shattering of larger grains does not require the unbinding of every molecule in the grain, while vaporization, in principle, does (or nearly does). It would be interesting to calculate more accurate binding energies for our SiO$_2$ and ADSil model grains to further validate this equation, and our results suggest it provides a reasonable first-guess approximation for the vaporization threshold velocity in grains larger than our adopted minimum grain size, but small enough that they do not agree with the continuum-dynamics predictions of \citet{Tielens1994ApJ...431..321T}, which we discuss further below.

\subsection{Shattering and Vaporization Mass Fractions as a Function of Velocity: Theory vs. Data}\label{subsec:discussion_mass_fractions}

\citet{Tielens1994ApJ...431..321T} go further \cje{than calculating shattering and vaporization threshold velocities} and derive the fraction of grain mass shocked to a pressure $P_1$, based on the dynamics of \cje{hemispherical} blast waves in solids, as a function of impact velocity $v_r = \mathscr{M}_r c_0$ in their Eq. 2.25:

\begin{equation}\label{eq:fMP_T1994}
\frac{M}{M_P} = \frac{(1 + 2\mathscr{R})}{2(1 + \mathscr{R})^{16/9}}\sigma_{r}^{-1/9}\left(\frac{\mathscr{M}_r^2}{\sigma_1\mathscr{M}_1^2}\right)^{8/9}
\end{equation}

\noindent where $M_P$ is the ``projectile'' (i.e. impactor) grain mass and

\begin{equation}
\sigma(\mathscr{M}) \equiv \frac{0.3(s + \mathscr{M}^{-1} - 0.11)^{1.3}}{s + \mathscr{M}^{-1} - 1}
\end{equation}

\noindent so that $\sigma_r = \sigma(\mathscr{M}_r / (1 + \mathscr{R}))$ and $\sigma_1 = \sigma(\mathscr{M}_1)$ are order-unity factors normalizing the energy associated with the self-similar blast wave in the grain. Note that this fraction is related to a specified internal grain pressure $P_1$ through the term $\mathscr{M}_1\equiv\mathscr{M}_{r1}/(1 + \mathscr{R})$ where $\mathscr{M}_{r1}$ is defined by substituting for $\mathscr{M}_r$ in Eq. \ref{eq:Tielens1994_2_16}. We note that Eq. \ref{eq:fMP_T1994} is derived for the late-stage of a non-energy-conserving hemispherical blast wave in a uniform, continuous plane-parallel medium. It is therefore, in principle, not appropriate for the collision between two grains of similar size, as Eq. \ref{eq:fMP_T1994} evaluates to $2.4$ at the threshold shattering velocity for silicate material properties, when physically this value should be bounded above by unity. 

\begin{figure}
    \centering
    \includegraphics[width=\linewidth]{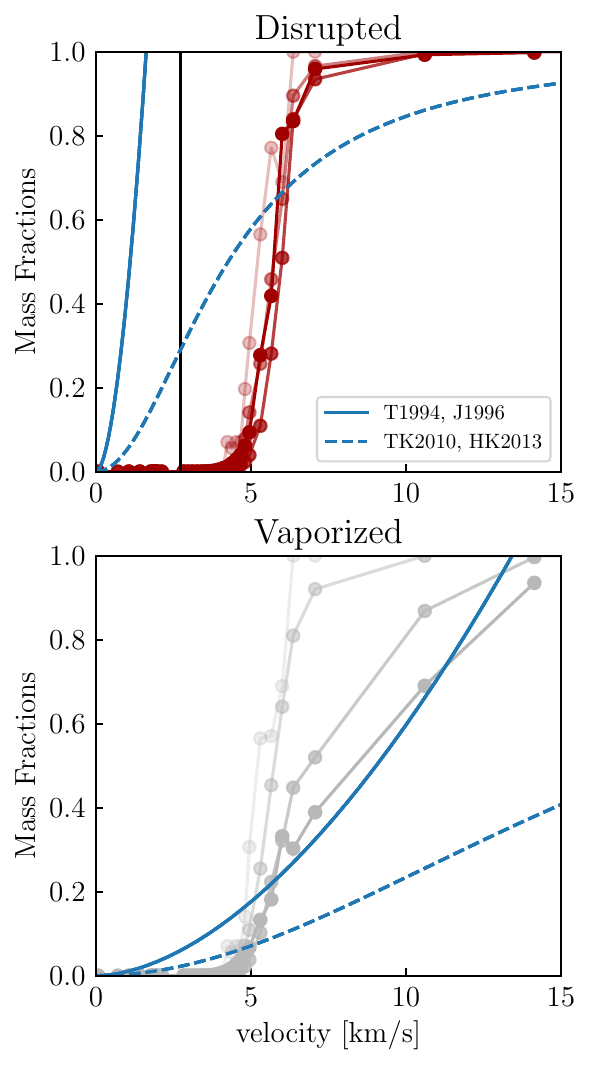}
    \caption{Disrupted and Vaporized Fractions as a function of collision velocity compared to models for SiO$_2$. \cje{Blue lines and the vertical black line indicate theoretical models and predicted thresholds, respectively.} Line transparency indicates grain size -- data from simulations of the \cje{largest} grains are the \cje{least transparent}.}
    \label{fig:SiO2_model_comp}
\end{figure}

\begin{figure}
    \centering
    \includegraphics[width=\linewidth]{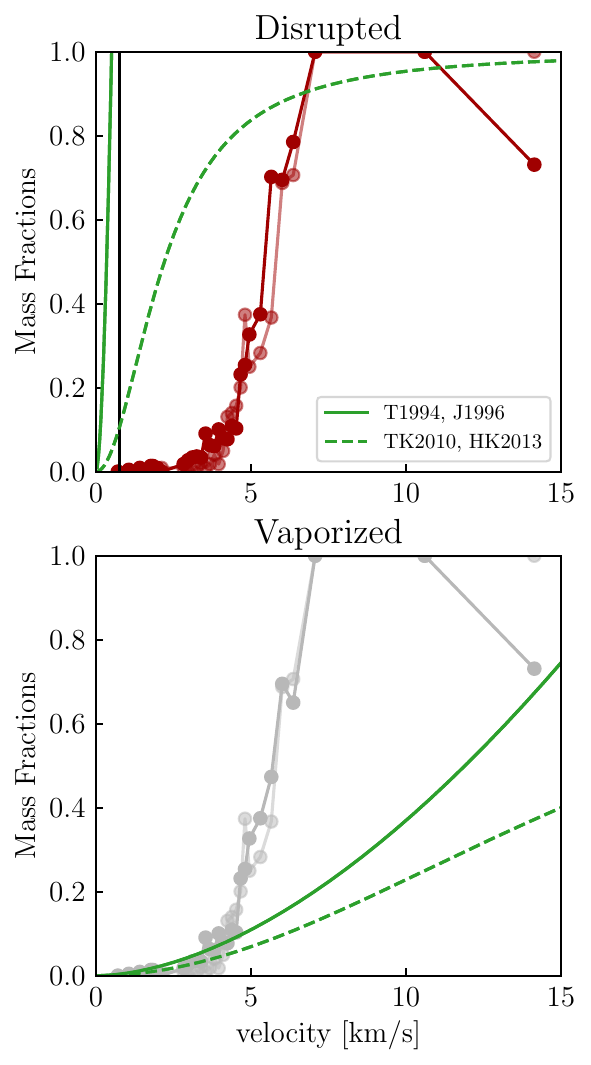}
    \caption{Same as Figure \ref{fig:SiO2_model_comp} but for ADSil.}
    \label{fig:AD_model_comp}
\end{figure}

An alternate formulation that has become popular in this kind of modeling was presented in \citet{KobayashiTanaka2010Icar..206..735K} and expanded upon in \citet{Hirashita2013EP&S...65.1083H}. In their ansatz the ejected mass from target grain $m_1$ impacted by the projectile grain of mass $m_2$ is given by

\begin{equation}\label{eq:M_ej_KT10}
M_{\rm ej} = \frac{\phi}{1 + \phi}m_1
\end{equation}

\noindent where

\begin{equation}
\phi = \frac{E_{\rm imp}}{m_1Q_D^*}
\end{equation}

\noindent and 

\begin{equation}
E_{\rm imp} = \frac{1}{2}\frac{m_1 m_2}{m_1 + m_2}v^2  
\end{equation}

\noindent and $Q_D^*$ is defined as the specific impact energy (i.e. energy per unit mass) at which $M_{\rm ej} = m_1/2$. \citet{Hirashita2013EP&S...65.1083H} compare to \citet{Jones1996ApJ...469..740J} by asserting $Q_D^* = P_1/(2 \rho_0)$ where $P_1$ is the threshold pressure for shattering and $\rho_0$ is the equilibrium grain density. This expression for $Q_D^*$, assuming $P_1$ is the shear modulus of \cje{the} material as done in \citet{Tielens1994ApJ...431..321T}, predicts a characteristic velocity set by $v = \sqrt{2Q_D^*} \approx 3$ km s$^{-1}$ (in the limit $m_2 \gg m_1$, which is appropriate for this comparison) close to the shattering velocity threshold obtained in \citet{Jones1996ApJ...469..740J}. This is compared to our simulation data in Figures \ref{fig:SiO2_model_comp} and \ref{fig:AD_model_comp}. 

From Figures \ref{fig:SiO2_model_comp} and \ref{fig:AD_model_comp}, it appears that for the small grains we study, neither the \citet{Tielens1994ApJ...431..321T} prescription nor the \citet{KobayashiTanaka2010Icar..206..735K, Hirashita2013EP&S...65.1083H} prescriptions convincingly describe the shattered or vaporized mass fractions as a function of collision velocity, \cje{except for possibly the \citet{Tielens1994ApJ...431..321T} model for vaporization applied to the largest grains we simulate (the darkest gray line in the bottom panel of Figure \ref{fig:SiO2_model_comp}). This suggests that the continuum dynamical predictions, at least regarding the process of vaporization, become approximately appropriate at grain sizes of $a\gtrsim 50$~\AA, but a conclusive interpretation of this agreement will require more simulations of similarly massive grains and further analysis, which we consider beyond the scope of the current paper.}

\subsection{Size Distributions: Differences with Previous Estimates}

The \citet{Jones1996ApJ...469..740J} prediction for shattered grain size distributions -- power-law with a velocity and grain-size independent slope of $\alpha = 3.3$ -- is not realized in our data (see Figure \ref{fig:s4s4_size_distributions}). This is unsurprising, given the plane-parallel geometry assumed for their calculation, and the steady-state fluid dynamical treatment of the grain cratering process. Instead, we see in the data that a single power-law fails to describe most remnant size or mass distributions, the shape of which is velocity dependent and, in the case of grain size, multimodal. This multimodality in the sizes, but not the masses, may be due to phase transitions in the grain material caused by the collisions which results in remnant grains with a range of densities, or changes in shattered product grain shapes. It may also be an artifact of the grain size calculation.

\subsection{Astrophysical and Observational Implications}

The new, higher velocity thresholds we obtain suggest that \cje{the smallest} interstellar dust \cje{grains} composed of silicate material \cje{may be} more resilient to shattering and vaporization than previously assumed. However, it is difficult to identify the precise effect of this difference on the numerous subsequent findings in which the \citet{Tielens1994ApJ...431..321T} and \citet{Jones1996ApJ...469..740J} results have been used, because interstellar grain populations are determined by the complicated interplay of many physical processes \cje{and the importance of small grains depends on the physical process or observable quantity being considered}. We can therefore only speculate on some of the most recent and most important applications of these values and discuss their potential implications.

The velocity distribution of, and therefore the distribution of pair-wise relative velocities between, dust grains in the interstellar medium is the result of complicated processes including charged-grain acceleration in magnetohydrodynamic turbulence (probably primary), gas-grain drag (also likely primary), ambipolar diffusion (probably negligible), Brownian motion (also probably negligible), and radiation pressure acceleration (negligible in \cje{some environments, dominant in others}) the theory of which is far from well-established \citep[see][for a recent review]{Ivlev2024-tn}. The standard reference on the subject is \citet{Yan2004-hp} who predict the scaling of dust grain velocities with grain size (which they assume determines a unique grain charge) in compressible magnetohydrodynamic turbulence, accounting for both fluid-dynamical drag and gyroresonant accelerations due to the dynamics of charged grains in fluctuating magnetic fields, applied to the parameters relevant for the characteristic thermodynamic phases of the ISM. These velocities are then compared to the \citet{Jones1996ApJ...469..740J} thresholds to set upper-limits on grain sizes due to collisional shattering in the warm neutral medium and warm ionized medium (as the parameters for all other phases predict grain velocities much smaller than these thresholds for all grain sizes, in their model). 

Interestingly, the grain velocities that \citet{Yan2004-hp} estimate increase so rapidly with grain size near the relevant thresholds that, despite \cje{the factor of $\sim 2$ difference between their thresholds and ours for small grains}, our modified velocity thresholds would only increase their stated maximal grain sizes by $\lesssim\mathcal{O}(50\%)$. Their estimate therefore remains qualitatively unchanged. However, \citet{Hirashita2010MNRAS.404.1437H} shows that the velocity dependence in the \citet{Yan2004-hp} model depends on assumed gas density, at least in the warm ionized medium (WIM), which does range between $n_H \sim 0.1-10\;{\rm cm}^{-3}$ in observations \citep{Haffner2009RvMP...81..969H, Langer2021A&A...651A..59L}. At higher densities, where grain collisions will be more frequent, the velocity dependence on grain size is much shallower and the velocity difference between \citet{Jones1996ApJ...469..740J} and ours would increase the predicted maximum grain size by about an order-of-magnitude. The effect of this changed velocity threshold is therefore, unsurprisingly, coupled to the complicated physics of the interstellar medium and will require more sophisticated modeling to understand the impact on the actual grain size distribution in the real ISM. 

We also note that the \citet{Yan2004-hp} calculations necessarily make many assumptions about the nature of interstellar turbulence (particularly the way its characteristic velocities vary with spatial scale), and dust grains (particularly the way their charge scales with grain size), and confirmation from fully-coupled simulations of these processes -- especially those resolving the turbulent dissipation scale where these assumptions are most directly manifest -- would be helpful for more definitive conclusions. \citet{Moseley2023MNRAS.518.2825M, Moseley2025MNRAS.542.1011M, HopkinsLee2016MNRAS.456.4174H, LeeHopkins2017MNRAS.469.3532L, Hopkins2022MNRAS.517.1491H, Soliman2024ApJ...974..136S} are recent efforts in this direction. 

The \citet{Jones1996ApJ...469..740J} velocity thresholds combined with the grain velocity scalings of \citet{Yan2004-hp} have become a standard ingredient of models that attempt to explain dust observational properties -- primarily the opacity law -- by dynamically evolving the dust size distribution in models of the galactic interstellar medium of varying complexity. These range from ``one-zone'' or ``semi-analytic'' models that derive and numerically solve systems of ordinary differential equations for average quantities of the interstellar medium \citep[e.g.][]{Jones1994ApJ...433..797J, Jones1996ApJ...469..740J, Hirashita2009MNRAS.394.1061H, Hirashita2010MNRAS.407L..49H, Hirashita2013EP&S...65.1083H, Asanon2013MNRAS.432..637A}, models that take Lagrangian pathlines from three-dimensional fluid dynamical simulations of the interstellar medium and evolve the dust content along these paths in post-processing \citep[e.g.][]{HirashitaAoyama2019MNRAS.482.2555H, Huang2021MNRAS.501.1336H}, to 3D simulations that evolve the grain size distribution on-the-fly and fully coupled to the gas dynamics and other physics \citep[e.g.][]{McKinnon2018MNRAS.478.2851M, Aoyama2017MNRAS.466..105A, Gjergo2018MNRAS.479.2588G, Li2021MNRAS.507..548L, Parente2022MNRAS.515.2053P, Narayanan2023ApJ...951..100N, Dubois2024A&A...687A.240D}. 

To evolve the dust size distribution these models must solve some form of the ``population balance'' (or ``particle population dynamics'' or Smoluchowski) equation, which formulates the evolution of the size distribution due to source and sink terms from coagulation and shattering \citep{Smoluchowski1916ZPhy...17..557S, Nakagawa1981Icar...45..517N, Mizuno1988A&A...195..183M}. Many of these studies reduce computational cost and complexity by evolving only two size bins for the dust \citep[see][]{Hirashita2015MNRAS.447.2937H}. Regardless, they generally incorporate the effects of shattering by assuming that dust grains with relative velocities greater than the shattering threshold are shattered with an ejecta distribution of $dn/da \propto a^{-3.3}$ as per the \citet{Jones1996ApJ...469..740J} predictions. However, none of these simulations, which have at best resolutions of $\sim 10$ pc, are able to resolve the scales relevant for grain acceleration and dynamics in a turbulent, magnetized, multiphase medium, a scale which is itself uncertain but likely comparable to the gyroradius of the grains around interstellar magnetic fields $r \lesssim 100~{\rm AU} \sim 10^{-4}~{\rm pc}$ \citep{Yan2004-hp, Slavin2012ApJ...760...46S} or at most the dissipation scale of interstellar turbulence $r \lesssim 4\times10^3~{\rm AU} \sim 2\times10^{-2}{\rm pc}$ \citep{Pindea2024A&A...690L...5P}. Thus they typically assume the \citet{Yan2004-hp} velocity scaling, in which case the effect of our \cje{different threshold shattering velocity} value would depend on the extent to which the \citet{Yan2004-hp} density dependence is included in the simulation prescription and which ISM densities contribute the most to grain size redistribution through collisions. 

Some other modeling efforts have instead used sub-grid models for grain velocity distributions determined by drag from interstellar gas velocity distributions based on a turbulent cascade \citep{HirashitaAoyama2019MNRAS.482.2555H, Li2021MNRAS.507..548L, Narayanan2023ApJ...951..100N}. \citet{Narayanan_2026} use this prescription in their simulations which predict that the size distributions of high-redshift galaxies are significantly ``flattened'' i.e. redistributing grain mass from high to low grain sizes from shattering in high velocity-dispersion gas, {\it only after} $z\approx 10$ (when the universe was $\sim 500$ Myr old). This provides a natural explanation for surprisingly blue and massive high-redshift ($z > 10$) galaxies observed with JWST \citep[e.g.][]{ArrabalHaro2023ApJ...951L..22A, CurtisLake2023NatAs...7..622C, Austin2023ApJ...952L...7A, Bunker2024A&A...690A.288B, Cullen2024MNRAS.531..997C, Morales2024ApJ...964L..24M, Topping2024MNRAS.529.4087T, RobertsBorsani2025ApJ...983...18R}, whose blue colors are, in these models, the consequence of flat dust opacity curves resulting from top-heavy grain size distributions (because shattering has not yet become important). \citet{Narayanan_2026} themselves note that the assumed grain-shattering thresholds are uncertain, an uncertainty the effects of which they plan to explore, and it will be interesting to see how their results depend on these assumptions. 

Regarding the predicted shattered grain size distributions, the direct observational implications for grains of the sizes we simulate are likely insignificant because all of these grains have sizes much smaller than the wavelengths of UV light, and are therefore well within the Rayleigh regime of electromagnetic absorption and scattering. In this regime grain cross sections are proportional to their total volume \citep{Galliano2022HabT.........1G}, of which larger grains are typically the dominant contribution \citep[see Figure 3 of][]{Hensley2023ApJ...948...55H}. Grain surface area is, however, dominated by small grains in the usual $\alpha > 3$ scaling for a power law $dn/da \propto a^{-\alpha}$ size distribution (which observations typically prefer), in which case surface-area-dependent processes such as grain growth via gas-phase accretion might be sensitive to the assumptions of grain shattering product distributions. Given the importance of grain growth via gas-phase accretion in setting the cosmic dust content throughout the history of the universe \citep[e.g.][]{Dwek1998ApJ...501..643D, Hirashita2000PASJ...52..585H, Feldmann2015MNRAS.449.3274F, Esmerian2022ApJ...940...74E, Esmerian2024ApJ...968..113E}, this effect could potentially be significant for interstellar grain evolution. The interplay of these effects requires further investigation in a comprehensive model of grain evolution in the ISM in which these processes are fully coupled. It will also be interesting to further investigate the shattered product size distribution as a function of grain size for larger grains than we study here, to understand the transition to the continuum dynamics results of \citet{Jones1996ApJ...469..740J}.

Our updated shattering thresholds are also relevant to the survival of dust grains in shocks caused by supernovae that propagate through the interstellar medium. \citet{Priestley2022MNRAS.516.2314P} analyzed the dust content of 7 galactic supernova remnants, placing low upper-limits on the mass fraction in small ($a\lesssim 10$ nm) dust grains in the hot ($T \gtrsim 10^6$ K), low-density ($n\sim 1$ cm$^{-3}$), post-shock gas. This result contradicts model predictions, which suggest that the small-size tail of the grain size distribution should be populated by grain shattering that occurs as grains pass through and are accelerated by the blast wave \citep{Kirchschlager2019MNRAS.489.4465K}. This tension between theory and observation may be alleviated with the higher shattering thresholds we predict in the present paper, because of which grain collisions will be less effective at redistributing dust mass to smaller grain sizes. However, as \citet{Priestley2022MNRAS.516.2314P} note, thermal sputtering should be effective in the post-shock hot phase, though given the relatively young age of the supernova remnants studied (see their Table 1), and the expected thermal sputtering timescales \citep{DraineSalpeter1979ApJ...231...77D, Draine2011piim.book.....D} this is likely not an entirely satisfactory explanation for the observed paucity of small grains. In practice these effects will work simultaneously, as was shown in \citet{Kirchschlager2022MNRAS.509.3218K}, where the effect of grain shattering increased the predicted dust destruction efficiency of supernova remnant shocks by an order-of-magnitude. Further analyses that use our updated grain shattering thresholds to re-evaluate the evolution of dust in interstellar shocks would therefore be interesting. 

\subsection{Prescription for Dust Models}

Based on the results of our MD simulations and the considerations discussed in the previous sections, we can formulate \cje{some} components of a prescription for the inclusion of our MD simulation results in simulations of the interstellar grain population on temporal and spatial scales interesting for astrophysics: \cje{for grains with size $a \lesssim 50$~\AA,} we recommend adopting \cje{the higher shattering thresholds given in Table~\ref{table:threshold_velocities}}, for which a step function in the disrupted mass fraction as a function of velocity is justified by the rapid transition shown in Figures \ref{fig:SiO2_model_comp} and \ref{fig:AD_model_comp}.

Above the shattering velocity thresholds, a prescription for the mass fraction shattered vs. vaporized for grains of radius $a \gtrsim 30$~\AA, is less obvious from our simulations, given the grain-mass-dependent behavior of the vaporized fraction most clearly shown in the bottom panel of Figure \ref{fig:SiO2_model_comp}. Visually, it appears the vaporized fraction could potentially be approaching the \citet{Tielens1994ApJ...431..321T} scaling (Eq. \ref{eq:fMP_T1994}) as grain size increases -- this is plausible, as the \citet{Tielens1994ApJ...431..321T} expression comes from a continuum dynamics derivation which is better justified for increasingly larger grains. It would be interesting to pursue a derivation of the \citet{Tielens1994ApJ...431..321T} equation modified to account for the effects of small grain size (possibly the relatively stronger surface free energy compared to larger grains), and/or modified to model collisions between equal-size grains (as opposed to a plane-parallel geometry), but we consider these goals beyond the scope of the present paper. We therefore cannot with our simulation data confidently give a prescription for vaporized fraction as a function of grain velocity that is applicable to all grain sizes, but we aim to extend our theoretical and numerical efforts towards this goal in the future.

Our results (Figures \ref{fig:SiO2_hetero_f_shat_vap} and \ref{fig:AD_hetero_f_shat_vap}) also suggest that shattering and vaporization need only be considered for grains of comparable (i.e. within a factor of 10) mass, because collisions between grains with larger mass ratios do not significantly disrupt the larger grain. More sophisticated prescriptions would need justification from a broader statistical sample of grain materials and collision parameters which we leave to future work. 

\section{Conclusions}\label{sec:conclusion}

We have for the first time (to the knowledge of the authors) run molecular dynamics simulations of colliding silicate dust grains of radii $a \in [5,50]$~\AA{} to understand their dynamics and outcomes, and confront popular analytical models of  such collisions with atomistic numerical simulations. Our simulations include collisions between both amorphous SiO$_2$ grains and grains with the chemical composition of astrodust silicates suggested by \citet{draine2021dielectric}.  We find:

\begin{itemize}
    \item Grain shattering thresholds for the \cje{two silicate compositions} we simulate are $\sim$ 6 km/s, a factor of approximately 2 higher than the canonical value of 2.7 km/s from \citet{Jones1996ApJ...469..740J}. \cje{We speculate that the physical origin of this discrepancy is some combination of additional physical effects relevant to the small grain sizes considered here -- namely the continuum dynamical assumption of the \citet{Tielens1994ApJ...431..321T} derivation and the increased importance of surface free energy -- but we lack a quantitative physical explanation.}
    \item The size distributions of shattered products are generally not well-described by power law distributions as predicted by \citet{Jones1996ApJ...469..740J} though this is not surprising because our grains are too small to satisfy the assumptions of the \citet{Jones1996ApJ...469..740J} derivation. 
    \item The prescription from \citet{Tielens1994ApJ...431..321T} for the fraction of grain mass shocked to a given pressure (Eq. \ref{eq:fMP_T1994}) as a function of collision velocity \cje{does not agree with} the fraction of shattered or vaporized material observed in the numerical simulations \cje{with the possible exception of the largest grain size we simulate, for which the \citet{Tielens1994ApJ...431..321T} continuum predictions appear to be a good approximation (Figure~\ref{fig:SiO2_model_comp}), but confirmation of this agreement will require more computationally intensive simulations of large grains.}
    \item The model of \citet{Hirashita2013EP&S...65.1083H} for the same quantities \cje{also does not} match the simulations.
    \item We provide updated shattering velocity thresholds for \cje{two} candidate \cje{silicate} grain materials \cje{(Table~\ref{table:threshold_velocities})} for use in future models of the dust population evolution on astrophysical scales. \cje{Our calculations predict that} shattering and vaporization only need to be considered for collisions between grains with mass ratios less than 10.
    \item Further simulations with a larger range of grain sizes and collision parameters, as well as grain material properties, will be needed to \cje{fully} understand how grain collision outcomes transition to the \citet{Tielens1994ApJ...431..321T} results for grains large enough to be modeled with continuum dynamics.
\end{itemize}

\begin{acknowledgements}
Financial support from the Knut and Alice Wallenberg foundation is gratefully acknowledged through grant nr. KAW 2020.0081. The computations were enabled by resources provided by the National Academic Infrastructure for Supercomputing in Sweden (NAISS), partially funded by the Swedish Research Council through grant agreement no. 2022-06725. \cje{BL acknowledges funding from the European Union’s Horizon Europe research and innovation programme under the Marie Sk\l{}odowska-Curie grant agreement No. 101126636.} CJE thanks Mike Shull and Felix Priestley for alerting him to relevant references.
\end{acknowledgements}

%
\bibliographystyle{aa} 
\bibliography{ref} 
%
\begin{appendix} 
\section{\cje{A simple re-derivation of the} \citet{Jones1996ApJ...469..740J} and \citet{Tielens1994ApJ...431..321T} Threshold Velocities Calculation}
\label{appendix:literature}

\citet{Jones1996ApJ...469..740J} uses the results of \citet{Tielens1994ApJ...431..321T} to calculate the minimum velocity for shattering (specifically crater formation, in their continuum dynamical treatment) which they posit corresponds to ``an initial shock pressure [in the target colliding grain] equal to the critical pressure of the target material,'' where they are considering the case of a spherical grain colliding with a much larger grain represented by a plane-parallel surface. This initial shock pressure is derived from the strong shock jump conditions \citep{McKeeHollenbach1980ARA&A..18..219M}, and is stated in \citet{Tielens1994ApJ...431..321T} equation 2.16 as 

\begin{equation}\label{eq:Tielens1994_2_16_appendix}
    \frac{P_{1i}}{\rho_0 c_0^2} = \left(\frac{\gamma + 1}{2}\right)\frac{\mathscr{M}_r^2}{(1 + \mathscr{R})^2} = \left(s + \frac{1 + \mathscr{R}}{\mathscr{M}_r}\right)\frac{\mathscr{M}_r^2}{(1 + \mathscr{R})^2}
\end{equation}

\noindent where $P_{1i}$ is the target grain initial shock pressure, $\rho_0$ is the target material density, $c_0$ is the target material sound speed, $\gamma$ is the adiabatic index (i.e. ratio of specific heats) which defines the target material equation of state, $\mathscr{M}_r \equiv v_r/c_0$ is the impact velocity $v_r$ Mach number in the target, $\mathscr{R} \equiv \sqrt{\frac{s\rho_0}{s_P \rho_{0P}}}$ is a function of the target and projectile (subscript $P$) material properties where $s$ is defined by an empirically calibrated relationship between the shock velocity and the velocity of the shocked matter $v_s = c_0 + s v_1$. \cje{Here we present an algebraically simpler derivation than appears in that paper, for the benefit of future readership.}

\cje{Using the material sound speed, $c_0$, to write the pressure and velocities in dimensionless form,}

\begin{equation}
\phi_1 \equiv \frac{P_1}{\rho_0 c_0^2},\;\;\;\mathscr{M}_1 \equiv \frac{v_1}{c_0},\;\;\;\mathscr{M}_s \equiv \frac{v_s}{c_0}
\end{equation}

\noindent \cje{the jump-shock condition, eq. 2.4 in \citet{Tielens1994ApJ...431..321T},}

\begin{equation}
    P_1 = \rho_0 v_1 v_s
\end{equation}

\noindent \cje{gives}

\begin{equation}\label{equ:jump_shock_simple}
    \phi_1 = \mathscr{M}_1\mathscr{M}_s
\end{equation}

\noindent \cje{Combine this with the empirical relationship between the solid shock velocity $v_s$ and the velocity of the shocked matter $v_1$ \citep[eq. 2.8 in][]{Tielens1994ApJ...431..321T, MCQUEEN1970293}}

\begin{equation}
\mathscr{M}_s = 1 + s\mathscr{M}_1
\end{equation}

\noindent \cje{and with eq.~\ref{equ:jump_shock_simple} we have}

\begin{equation}
    \phi_1 = \mathscr{M}_1(1 + s\mathscr{M}_1).
\end{equation}

\noindent  \cje{For same-material collisions, symmetry gives $v_1 = v_r/2$, and more generally $v_1 = v_r/(1 + \mathscr{R})$ \citep[eq. 2.14 in][]{Tielens1994ApJ...431..321T}, so}

\begin{equation}
    \phi_1 = \frac{\mathscr{M}_r}{1 + \mathscr{R}}+ s\frac{\mathscr{M}_r^2}{(1 + \mathscr{R})^2}
\end{equation}

\noindent \cje{thereby recovering eq. \ref{eq:Tielens1994_2_16_appendix} \citep[i.e. eq. 2.16 in][]{Tielens1994ApJ...431..321T} for which the solution is}

\begin{equation}
    \mathscr{M}_r = \frac{2(1 + \mathscr{R})\phi_1}{1 + \sqrt{1 + 4 s \phi_1}}
\end{equation}

\noindent \cje{i.e. equation 2.32 in \citet{Tielens1994ApJ...431..321T} and A2 in \citet{Jones1996ApJ...469..740J}, where $\mathscr{M}_r/(1 + \mathscr{R})$ has been substituted for $\mathscr{M}_1$.}

\end{appendix}

\end{document}